\PassOptionsToClass{cleveref}{jmlr}
\documentclass{colt2026}
\usepackage{mathtools,microtype,needspace,booktabs,pifont,array, times, enumitem}
\usepackage{thm-restate}
\addtotheorempostheadhook[proof]{\ }

\allowdisplaybreaks[1]
\numberwithin{theorem}{section}
\aliascntresetthe{lemma}
\aliascntresetthe{proposition}
\crefname{lemma}{Lemma}{Lemmas}
\Crefname{lemma}{Lemma}{Lemmas}
\crefname{theorem}{Theorem}{Theorems}
\Crefname{theorem}{Theorem}{Theorems}

\newcommand{\Reg}{R}
\DeclareMathOperator{\KL}{KL}
\DeclareMathOperator{\Var}{Var}
\DeclareMathOperator{\Cov}{Cov}
\DeclareMathOperator{\softmax}{softmax}
\newcommand{\one}{\mathbf 1}
\newcommand{\E}{\mathbb E}
\newcommand{\R}{\mathbb R}
\newcommand{\ip}[2]{\langle #1,#2\rangle}
\newcommand{\norm}[1]{\lVert #1\rVert}
\newcommand{\tableYes}{\textcolor[rgb]{0,0.8,0}{\ding{51}}}
\newcommand{\tableNo}{\textcolor[rgb]{1,0,0}{\ding{55}}}
\newcommand{\tableUnknown}{\textcolor[rgb]{0.6,0.47,0}{\textbf{?}}}
\makeatletter
\unless\ifanonsubmission
  \unless\iffinalsubmission
    \g@addto@macro\ps@jmlrtps{%
      \def\@oddhead{\scriptsize Preprint\hfill}%
      \let\@evenhead\@oddhead
    }
  \fi
\fi
\makeatother
\title[Cautiously Adaptive Optimistic Hedge]{%
  \resizebox{\linewidth}{!}{\shortstack{%
    Cautiously Adaptive Optimistic Hedge in General-Sum Games:\\
    Constant Regret and (Doubly) Exponential Improvement}}}
\coltauthor{\Name{Junsoo Ha}\Email{junsoo.ha.contact@gmail.com}
\AND \Name{Haipeng Luo}\Email{haipengl@usc.edu}\\
\addr University of Southern California}
\hypersetup{pdfauthor={Junsoo Ha, Haipeng Luo},pdftitle={Cautiously Adaptive Optimistic Hedge in General-Sum Games: Constant Regret and (Doubly) Exponential Improvement},pdfsubject={Preprint; adaptive step sizes and individual regret}}
\begin{document}
\maketitle
\begin{abstract}
Recent work established logarithmic~\citep{ha_log_hedge_2026} and even horizon-independent~\citep{ha_horizon_hedge_2026} external regret bounds for Optimistic Hedge with a constant step size in general-sum games.
However, the horizon-independent upper bound of \citet{ha_horizon_hedge_2026} left its dependence on the number of players $n$ and the number of actions $d$ implicit, and \citet{dagan_eisen_erez_koren_2026} recently showed that such dependence can be exponential, namely $2^{\Omega(nd)}$.
In this work, we ask whether simple adaptive step sizes can overcome such a pathological lower bound.
We answer this question affirmatively with \emph{Cautiously Adaptive Optimistic Hedge} (CAOH), which chooses its step sizes following the principle of \emph{cautiousness}~\citep{soleymani_cautious_2025} and deliberately slows down a player when they perform excessively well.
We show that CAOH achieves a horizon-independent $O(\sqrt n\log d)$ individual regret bound under self-play, demonstrating an exponential improvement in the number of players $n$ and a doubly exponential improvement in the number of actions $d$ compared to Optimistic Hedge with a constant step size.
This upper bound matches the state-of-the-art upper bound achieved by the MORM algorithm of~\citet{soleymani_piliouras_morm}.
Unlike MORM, however, CAOH additionally admits an efficient kernelized implementation for extensive-form games, and more generally, $0/1$-polyhedral games whose normal-form representations can be exponentially large.
\end{abstract}

\section{Introduction}\label{sec:introduction}
In strategic games, each player faces a loss sequence generated by the other players' choices~\citep{fudenberg_levine_1998}.
In such settings, no-regret learning provides a decentralized route to an equilibrium of the game~\citep{cesa_bianchi_lugosi_2006}.
For instance, external regret compares each player's cumulative loss with that of its best fixed action in hindsight, and when every player has sublinear external regret, their time-averaged joint play approaches the set of coarse correlated equilibria (CCE)~\citep{moulin_vial_1978}.

While the minimax regret against the worst-case adversaries follows the standard rate $\Theta(\sqrt T)$ for a horizon $T$~\citep{freund_schapire_1997,cesa_bianchi_et_al_1997}, self-play dynamics can induce more predictable losses as all players share the same learning rule, hence, faster convergence.
For example, Optimistic Hedge exploits this predictability by using the most recent loss vector as its next prediction~\citep{rakhlin_sridharan_2013} to attain much smaller regret bounds~\citep{syrgkanis_rvu_2015,daskalakis_fishelson_golowich_2021}.
Across the long line of work on this topic which we will review in \Cref{sec:related-work}, recent progress has shown that Optimistic Hedge with a constant step size can attain logarithmic~\citep{ha_log_hedge_2026} and even horizon-independent external regret bounds~\citep{ha_horizon_hedge_2026}.
However, the horizon-independent upper bound of \citet{ha_horizon_hedge_2026} left its dependence on the number of players $n$ and the number of actions $d$ implicit, and \citet{dagan_eisen_erez_koren_2026} recently showed that such dependence can be exponential by establishing a $2^{\Omega(nd)}$ regret lower bound.
Noting that this lower bound crucially relies on the constant step size of Optimistic Hedge, we ask the following question:
\begin{center}
\itshape Can we circumvent the exponential lower bound with adaptive step sizes?
\end{center}

\paragraph{Contributions.}
We answer this question affirmatively with \emph{Cautiously Adaptive Optimistic Hedge} (CAOH), a simple variant of Optimistic Hedge with adaptive step sizes, and show that it achieves a constant $O(\sqrt n\log d)$ individual regret bound.
The significance of this result is three-fold:
\begin{itemize}[leftmargin=*]
    \item First, it shows that adaptive step sizes can lead to an exponential improvement in dependence on $n$, and a doubly exponential improvement in dependence on $d$, compared with constant step sizes~\citep{dagan_eisen_erez_koren_2026}.

    \item Second, it matches the state-of-the-art constant regret bound for this problem, which is achieved by Multiplicatively Optimistic Regret Matching (MORM)~\citep{soleymani_piliouras_morm}.

    \item Finally, our algorithm admits an efficient kernelized implementation for $0/1$-polyhedral games such as extensive-form games~\citep{farina_kernel_2022}, while it is unclear if MORM~\citep{soleymani_piliouras_morm} can be efficiently implemented for such settings.
\end{itemize}

Our adaptive step sizes are easy to compute and conceptually related to the idea of \emph{cautiousness}~\citep{soleymani_cautious_2025}: the algorithm reduces each player's step sizes as the magnitude of its \emph{soft regret}, i.e., a regret against an entropy-regularized fixed strategy, grows larger.
In particular, when a player performs \textit{too well} with very negative soft regret, the algorithm reduces the step size and deliberately slows down the player.
This somewhat counterintuitive idea is important to maintain the balance among players and an overall harmonic learning in games~\citep{soleymani_cautious_2025}.
While the algorithm of \citet[Section~3.3]{soleymani_cautious_ec_2025} selects its step sizes via scalar optimization, our algorithm computes them in a closed form without relying on iterative subroutines. %

\section{Related work}\label{sec:related-work}
\paragraph{No-regret learning in games.}
No-regret learning connects individual regret guarantees to equilibrium approximation.
When all players have sublinear external regret, time-averaged joint strategies yield approximate Nash equilibria through their marginals in two-player zero-sum games~\citep{freund_schapire_games_1999} and approximate CCE in general-sum games~\citep[Section~7.4]{cesa_bianchi_lugosi_2006}.
Correlated equilibrium (CE)~\citep{aumann_1974} strengthens the notion of equilibrium, and can be approximated by no-swap-regret algorithms~\citep{cesa_bianchi_lugosi_2006}.
The Blum--Mansour reduction~\citep{blum_mansour_2007} builds such algorithms from external-regret learners and underlies fast convergence to CE~\citep{anagnostides_log_swap_2022,tsuchiya_hybrid_2026,mai_swap_2026}.
In this work, we focus on individual external regret bounds and CCE convergence rates.
\Cref{tab:regret-comparison} compares our results and all existing $o(\sqrt{T})$ individual external regret bounds that we are aware of.

\begin{table}[!t]
\centering
\footnotesize
\setlength{\tabcolsep}{3pt}
\renewcommand{\arraystretch}{1.05}
\caption{Self-play individual external regret bounds over $T$ rounds in general-sum games with $n$ players and at most $d$ actions per player. OH denotes Optimistic Hedge with a constant step size.
``Single loop'' means closed-form updates without iterative subroutines.~``Kernelization'' denotes the existence of an efficient implementation via kernelization for $0/1$-polyhedral games~\citep{farina_kernel_2022}.
The worst-case lower bound of \cite{dagan_eisen_erez_koren_2026} applies for $n\ge3$, sufficiently large $d$, and sufficiently small constant step sizes.
The notation $O_{n,d}(1)$ for~\citet{ha_horizon_hedge_2026} denotes a bound that is constant when treating both $n$ and $d$ as constants.
}
\label{tab:regret-comparison}
\begin{tabular}{@{}>{\raggedright\arraybackslash}p{0.32\linewidth}>{\raggedright\arraybackslash}p{0.18\linewidth}>{\raggedright\arraybackslash}p{0.185\linewidth}cc@{}}
\toprule
Reference & Algorithm & Regret Bound & Single loop & Kernelization \\
\midrule
\citet{syrgkanis_rvu_2015} & OFTRL/OMD & $O(\sqrt n\log d\,T^{1/4})$ & \tableYes & \tableUnknown \\
\citet{chen_peng_2020} & OH & $O((\log d)^{5/6}T^{1/6})$ & \tableYes & \tableYes \\
\citet{daskalakis_fishelson_golowich_2021} & OH & $O(n\log d\log^4 T)$ & \tableYes & \tableYes \\
\citet{farina_convex_2022} & LRL-OFTRL & $O(nd\log T)$ & \tableNo & \tableUnknown \\
\citet{soleymani_cautious_2025,soleymani_cautious_ec_2025} & COMWU & $O(n\log^2 d\log T)$ & \tableNo & \tableYes \\
\citet{liu_farina_ozdaglar_2026} & ECHO-OFTRL & $O(n^{21}\log^4 d)$ & \tableNo & \tableYes \\
\citet{abbadi_laraki_mertikopoulos_2026} & HOOD & $O(n^3\log^2 d)$ & \tableNo & \tableYes \\
\citet{ha_log_hedge_2026} & OH & $O(\sqrt n\log d\log T)$ & \tableYes & \tableYes \\
\citet{ha_horizon_hedge_2026} & OH & $O_{n,d}(1)$ & \tableYes & \tableYes \\
\citet{dagan_eisen_erez_koren_2026} & OH & \textcolor{red}{\boldmath$2^{\Omega(nd)}$} & \tableYes & \tableYes \\
\citet{soleymani_piliouras_morm} & MORM & $O(\sqrt n\log d)$ & \tableYes & \tableUnknown \\
\textbf{This work} & \textbf{CAOH} & $O(\sqrt n\log d)$ & \tableYes & \tableYes \\
\bottomrule
\end{tabular}
\end{table}

\paragraph{Constant regret bounds.}
For many years, the dependence on time horizon $T$ in existing individual external regret bounds has remained at best polylogarithmic~\citep{daskalakis_fishelson_golowich_2021, farina_convex_2022, soleymani_cautious_2025,soleymani_cautious_ec_2025}, and whether a horizon-independent bound is possible was a major open question until very recently.
In particular, two concurrent results first answered this question affirmatively: both ECHO-OFTRL~\citep{liu_farina_ozdaglar_2026} and HOOD~\citep{abbadi_laraki_mertikopoulos_2026} attained horizon-independent external regret bounds by combining lifted entropic regularization with higher-order prediction.
In particular, ECHO-OFTRL~\citep{liu_farina_ozdaglar_2026} used a cascade of exponential moving averages, and HOOD~\citep{abbadi_laraki_mertikopoulos_2026} used a discounted high-order predictor.
Their strategy updates reduce to one-dimensional optimization, which can be solved by iterative subroutines.
Shortly after these two results, \citet{soleymani_piliouras_morm} proposed the MORM algorithm and attained the state-of-the-art $O(\sqrt n\log d)$ regret bound via single-loop multiplicative optimism.
Our algorithm matches this bound while retaining the strategy update rule of Optimistic Hedge so that it enjoys an efficient kernelization for extensive-form games, or more generally, $0/1$-polyhedral games~\citep{farina_kernel_2022}.

\paragraph{Optimistic Hedge.}
Optimistic Hedge is arguably one of the most canonical algorithms for uncoupled learning in games. It uses the most recent loss vector as its prediction in entropic optimistic FTRL (OFTRL)~\citep{rakhlin_sridharan_2013}.
A long line of research on Optimistic Hedge with constant step sizes improved its individual regret bounds from polynomial to horizon-independent upper bounds.
For instance, through their Regret bounded by Variation in Utilities (RVU) framework, \citet{syrgkanis_rvu_2015} exploited self-play-induced stability to obtain an $O(\sqrt n\log d\,T^{1/4})$ individual regret bound in general-sum games.
\citet{chen_peng_2020} improved this bound to $O((\log d)^{5/6}T^{1/6})$ in two-player games, and \citet{daskalakis_fishelson_golowich_2021} exploited higher-order smoothness of self-play to further obtain a polylogarithmic $O(n\log d\log^4 T)$ upper bound for general $n$-player games.
More recently, \citet{ha_log_hedge_2026} sharpened this bound to a logarithmic $O(\sqrt n\log d\log T)$ upper bound, and \citet{ha_horizon_hedge_2026} further obtained a horizon-independent $O_{n,d}(1)$ regret bound via a nonconstructive finite-order relation, leaving its dependence on the game size ($n,d$) implicit.
\citet{dagan_eisen_erez_koren_2026} further investigated such dependence and showed that, somewhat surprisingly, Optimistic Hedge with constant step sizes can incur a pathological regret $2^{\Omega(nd)}$.
Our work introduces a simple fix via adaptive step sizes, and shows that it leads to an exponential improvement in the dependence on $n$ and a doubly exponential improvement in the dependence on $d$.

\paragraph{Cautious optimism.}
Intuitively, a large negative regret can be interpreted as an indication that the player is outperforming every fixed action, and therefore, has room to slow down its learning.
Cautious optimism~\citep{soleymani_cautious_2025,soleymani_cautious_ec_2025} uses this room to make self-play more predictable and improves individual regret guarantees.
This adaptive pacing is implicit in the lifted OFTRL in \citet{farina_convex_2022}, which achieves a $O(nd\log T)$ regret upper bound.
\citet{soleymani_cautious_2025,soleymani_cautious_ec_2025} formalized this adaptive pacing as an optimization problem, generalized the method to other regularizers, and obtained $O(n\log^2 d\log T)$ individual regret for entropic regularization.
Our algorithm CAOH implements the same principle through simple adaptive step sizes based on the magnitude of soft regret, without relying on iterative subroutines.

\section{Problem setup, algorithm, and main results}\label{sec:setup}

In this section, we formally introduce the problem setup, notations, algorithm, and main result.

\subsection{Setup and notation}
We study $n$-player general-sum games, where each player $i$ has $d_i\ge1$ actions and a loss function $\ell_i:\prod_{j\in [n]} [d_j]\to[0,1]$, where $[k]$ denotes the set $\{1,\ldots,k\}$ for any $k\in\mathbb{N}=\{1,2,\ldots\}$. Let $\Delta_k\coloneqq\{p\in\R_+^k: \|p\|_1=1\}$ be the $(k-1)$-dimensional simplex.
A distribution $x_i \in \Delta_{d_i}$ is called a mixed strategy for player $i$, and a joint profile of mixed strategies $(x_1, \ldots, x_n)\in \prod_{j\in [n]}\Delta_{d_j}$ for all players induces a loss vector $\ell_i(x_{-i})$ for each player $i$, where we define
\begin{equation}
 [\ell_i(x_{-i})]_{a_i}
 \coloneqq\sum_{a_{-i}\in \prod_{j\in[n]\setminus\{i\}} [d_j]}
 \ell_i(a_i,a_{-i})\prod_{j\in[n]\setminus\{i\}}x_{j,a_j}.
 \label{eq:game}
\end{equation}
At each round $t\ge1$, $n$ players simultaneously choose their mixed strategies $x_i^t\in\Delta_{d_i}$, and observe their loss vector $\ell_i^t=\ell_i(x_{-i}^t)$. With the cumulative loss vector $L_i^t=\sum_{s=1}^t\ell_i^s$, player $i$'s external regret $\Reg_i^T$ against the best fixed action in hindsight is
\begin{equation}
 \Reg_i^T\coloneqq\sum_{t=1}^T\ip{x_i^t}{\ell_i^t}-\min_{a\in[d_i]}L_{i,a}^T.
 \label{eq:regret}
\end{equation}
Let the maximum number of actions be $d=\max_{i\in[n]} d_i$. We write the KL divergence between two strategies $p,q\in\Delta_k$ as $\KL(p\|q)\coloneqq\sum_{a\in[k]} p_a\log(p_a/q_a)$, and define the map $\softmax:\R^k\to\operatorname{relint}(\Delta_k)$ by $[\softmax(v)]_a\coloneqq e^{v_a}/\sum_{b\in[k]} e^{v_b}$.

\subsection{Cautiously Adaptive Optimistic Hedge}\label{sec:caoh}
Standard Hedge~\citep{freund_schapire_1997} with a constant inverse step size $\beta > 0$ is a simple algorithm where every player plays $\softmax(-L_i^t/\beta)$ at time $t+1$.
Optimistic Hedge is a variant of Hedge where every player plays $\softmax(-(L_i^t+\ell_i^t)/\beta)$ at time $t+1$ instead~\citep{rakhlin_sridharan_2013}.
The intuition for double counting the most recent loss vector $\ell_i^t$ is that in slowly changing learning dynamics, $\ell_i^t$ can be a good prediction for the next loss vector $\ell_i^{t+1}$.
\begin{algorithm2e}[t]
\SetAlCapHSkip{0pt}
\caption{Cautiously Adaptive Optimistic Hedge (CAOH) for player $i$}
\label{alg:caoh}
\DontPrintSemicolon
\SetAlgoVlined
\SetKwInput{KwInit}{Initialize}
\KwIn{number of actions $d_i$, base inverse step size $\beta>0$, and scale $\Lambda_d=16(1+\log d)$}
\KwInit{$x_i^1\gets\one/d_i$, $\beta_i^1\gets\beta$, $L_i^0\gets0$, and $A_i^0\gets0$}
\For{$t=1,2,\ldots$}{
  Play $x_i^t$ and observe the loss vector $\ell_i^t$\;
  $L_i^t\gets L_i^{t-1}+\ell_i^t$,
  \quad $A_i^t\gets A_i^{t-1}+\ip{x_i^t}{\ell_i^t}$\;
  $\hat{R}_i^t\gets A_i^t+\beta_i^t\log\!\left(\frac{1}{d_i}
    \sum_{a=1}^{d_i}e^{-L_{i,a}^t/\beta_i^t}\right)$\;
  $\beta_i^{t+1}\gets\sqrt{\beta^2+(\hat{R}_i^t/\Lambda_d)^2}$\;
  $x_i^{t+1}\gets\softmax\!\left(-\frac{L_i^t+\ell_i^t}{\beta_i^{t+1}}\right)$\;
}
\end{algorithm2e}

Now, we are ready to present our algorithm \textit{Cautiously Adaptive Optimistic Hedge} (CAOH).
CAOH only changes the inverse step size of Optimistic Hedge from a fixed constant $\beta$ to a player-wise time-dependent adaptive inverse step size $\beta_i^{t+1}=\sqrt{\beta^2+(\hat{R}_i^t/\Lambda_d)^2}$.
Here, $\beta>0$ is a base inverse step size (to be chosen later), $\Lambda_d=16(1+\log d)$, and $\hat{R}_i^t = \sum_{s=1}^t \ip{x_i^s}{\ell_i^s} + \beta_i^t\log\!\left(\frac{1}{d_i}\sum_{a=1}^{d_i}e^{-L_{i,a}^t/\beta_i^t}\right)$ is a term we call \emph{soft regret}; see \algorithmref{alg:caoh} for details.
It is straightforward to verify that soft regret $\hat{R}_i^t$ has the following equivalent form
\begin{equation}
 \hat{R}_i^t=\sum_{s=1}^t\ip{x_i^s}{\ell_i^s}-\min_{p\in\Delta_{d_i}}
 \bigl\{\ip p{L_i^t}+\beta_i^t\KL(p\|\one/d_i)\bigr\}.
 \label{eq:soft-regret}
\end{equation}
In other words, soft regret $\hat{R}_i^t$ measures each player's cumulative loss (up to time $t$) relative to the best fixed strategy \textit{under entropic regularization}, and it recovers standard regret $\Reg_i^t$ when the entropic regularization is removed (that is, $\beta_i^t=0$).
Intuitively, the update rule in \algorithmref{alg:caoh} implements the \emph{cautiousness} principle similar to~\citep{soleymani_cautious_2025} by increasing the inverse step size as $\hat{R}_i^t$ becomes more negative, hence slowing down the player's learning when they perform excessively well and ensuring a balance among all players.

\paragraph{The importance of soft regret.}
We emphasize that using soft regret $\hat{R}_i^t$, instead of standard regret $R_i^t$, is crucial.
Indeed, in Appendix~\ref{app:ordinary-regret}, we prove a linear regret lower bound for the variant of CAOH that replaces soft regret $\widehat R_i^t$ with standard regret $R_i^t$.
The intuition is as follows. The softmax in the Optimistic Hedge strategy update rule assigns positive probability even to actions with large cumulative loss.
If we replace soft regret $\bar{R}_i^t$ in \algorithmref{alg:caoh} with standard regret $R_i^t$, losses from these actions can increase positive regret $R_i^t$ and thereby raise the next inverse step size $\beta_i^{t+1}$.
This creates a pathological feedback loop: the reduced step size sustains the losses that caused the step size reduction itself.
Soft regret breaks this loop by using a different comparator.
Unlike standard regret, soft regret uses the $\softmax(-L_i^t/\beta_i^t)$ strategy for its comparator, instead of a pure action.
This prevents the small probabilities assigned to actions with large cumulative loss causing unnecessary step size reductions, and allow CAOH to avoid the aforementioned pathological feedback loop.

\paragraph{Comparison with existing adaptive step sizes.}
There are several existing studies that also use adaptive step sizes for Optimistic Hedge; see for example~\citet{rakhlin_sridharan_2013, hsieh2021adaptive, tsuchiya2025corrupted}.
At a high level, their inverse step size (multiplied with $\log d$) is a certain \textit{regret upper bound}, such as those that depend on the cumulative loss variation.
On the other hand, our inverse step size (multiplied with $\Lambda_d = O(\log d)$) is directly the magnitude of the \textit{soft regret itself}, not an upper bound.
This not only enables us to implement the principle of cautiousness as discussed earlier, but also intuitively provides a more accurate characterization since regret upper bounds can be loose. 

\subsection{Main result}\label{sec:result}
\begin{figure}[t]
\centering
\includegraphics[width=\linewidth]{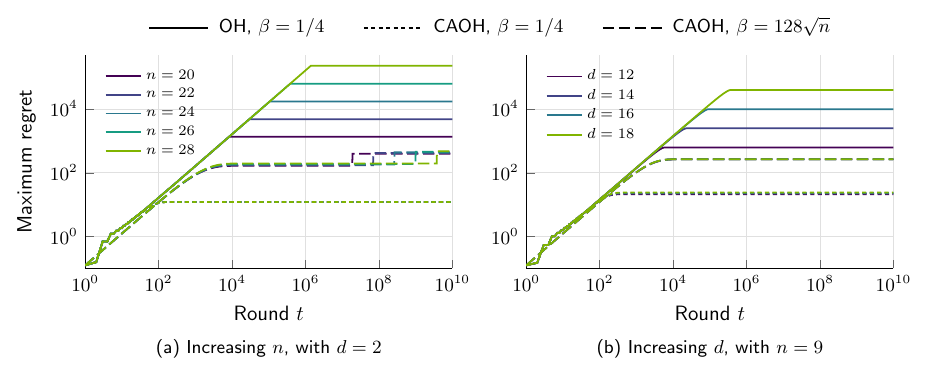}
\caption[Saturated regret as the game dimensions increase]{%
\textbf{\boldmath Maximum individual external regret over iterations.}
Solid and dashed curves represent Optimistic Hedge (OH) and Cautiously Adaptive Optimistic Hedge (CAOH), respectively. Colors distinguish values of $n$ in (a) and $d$ in (b). OH uses $\beta=1/4$, and short and long dashes show CAOH at $\beta=1/4$ and $\beta=128\sqrt n$, respectively. OH's saturated regret grows exponentially in both dimensions, while CAOH remains at significantly lower levels. This qualitative separation is consistent with the $2^{\Omega(nd)}$ lower bound for OH in \citet{dagan_eisen_erez_koren_2026} and the upper bound for CAOH in \cref{thm:main}. We defer the experiment details to Appendix~\ref{app:numerical-experiment}.}
\label{fig:dimension-sweeps}
\end{figure}

Our main result shows that the simple adaptive step sizes in \algorithmref{alg:caoh} suffice for Optimistic Hedge to attain a state-of-the-art constant regret in general-sum games under self-play.
\begin{theorem}[Constant regret bound under self-play]\label{thm:main}
If every player uses \algorithmref{alg:caoh} with $\beta=128\sqrt n$, then, for every player $i$ and horizon $T$, their external regret is bounded as
\begin{equation}
 \Reg_i^T\le5120\sqrt n\log d.
 \label{eq:selfplay}
\end{equation}
\end{theorem}
\noindent\Cref{thm:main} shows that the adaptive step sizes in \algorithmref{alg:caoh} suffice to overcome the exponential $2^{\Omega(nd)}$ regret lower bound that constant step sizes can suffer from~\citep[Theorem~6.24]{dagan_eisen_erez_koren_2026}.
We illustrate this (doubly) exponential separation of CAOH from Optimistic Hedge with constant step sizes in Figure~\ref{fig:dimension-sweeps}.
The resulting $O(\sqrt n\log d)$ individual regret matches the state-of-the-art upper bound recently achieved by MORM~\citep{soleymani_piliouras_morm} (\textit{c.f.} \Cref{tab:regret-comparison}).
Unlike MORM, however, CAOH retains the entropic update of Optimistic Hedge and admits an efficient kernelized implementation in extensive-form games or more generally, $0/1$-polyhedral games whose normal-form representations can have an exponential size~\citep{farina_kernel_2022}.\footnote{We note that the extra step of CAOH for computing soft regret is also kernelizable, since the key term $\sum_{a=1}^{d_i}e^{-L_{i,a}^t/\beta_i^t}$ is exactly the normalization factor of $\softmax\left(-L_i^{t}/\beta_i^t\right)$, which is kernelizable~\citep[Theorem~4.1]{farina_kernel_2022}.}
Finally, CAOH also enjoys an adversarial regret upper bound: in Appendix~\ref{app:adversarial}, we show that the individual external regret of a player using CAOH can be bounded as $O(\sqrt{T\log d})$ no matter how other players behave.

\section{Analysis}\label{sec:analysis}
In this section, we provide the main ideas behind our proof, with most details deferred to the appendix.
For convenience, we set $\ell_i^0=0$, $\beta_i^0=\beta_i^1=\beta$, and $\hat{R}_i^0=0$ for every player $i$ in the analysis.
These conventions extend the strategy formula to the first round and make the initial change in inverse step size (that is, $|\beta_i^1 - \beta_i^0|$) zero.

\subsection{From external regret to normalized soft regret}\label{sec:reduction}

We first point out that one common idea behind several existing results~\citep{farina_convex_2022, soleymani_cautious_2025, liu_farina_ozdaglar_2026, abbadi_laraki_mertikopoulos_2026} is to bound the positive part of the regret, that is, $\max\{\Reg_i^T,0\}$, by treating it as the standard regret in a lifted space.
Our analysis is somewhat similar, but instead of bounding $\max\{\Reg_i^T,0\}$, we bound $\max\{\hat{R}_i^T,0\}$, that is, the positive part of the soft regret.
The advantage is that to control $\max\{\hat{R}_i^T,0\}$, it turns out that it suffices to bound the \textit{normalized soft regret}, defined as $\bar{R}_i^t\coloneqq\hat{R}_i^t/\beta_i^{t+1}$,
without considering a lifted space.
This idea is summarized in the following two lemmas, whose proofs are deferred to Appendix~\ref{app:preliminary-estimates}.

\begin{restatable}[Soft regret approximates external regret]{lemma}{regretcomparison}\label{lem:regret-comparison}
\setlength{\emergencystretch}{2em}
For player $i$, fix arbitrary strategies $\{x_i^t\}_{t\ge1}$, an arbitrary loss sequence $\{\ell_i^t\}_{t\ge1}\subseteq[0,1]^{d_i}$, and positive inverse step sizes $\{\beta_i^t\}_{t\ge1}$. Then, for every $t\ge1$, the soft regret in \eqref{eq:soft-regret} satisfies
\begin{equation}
 \hat{R}_i^t\le\Reg_i^t\le \hat{R}_i^t+\beta_i^t\log d_i.
 \label{eq:regret-comparator}
\end{equation}
Additionally, suppose $\beta_i^1=\beta\ge4$ and that $\beta_i^{t+1}=\sqrt{\beta^2+(\hat{R}_i^t/\Lambda_d)^2}$ holds for every $t\ge1$. Then, the following holds for every $t\ge1$:
\begin{equation}
 \Reg_i^t\le\frac{\beta\Lambda_d}{8}+\frac98\max\{\hat{R}_i^t,0\}.
 \label{eq:regret-target}
\end{equation}
\end{restatable}

\begin{restatable}[Normalized soft regret controls soft regret]{lemma}{normalizedsoft}\label{lem:normalized-to-soft}
Fix a player $i$ and an integer $t\ge1$. Suppose $\beta>0$ and $\beta_i^{t+1}=\sqrt{\beta^2+(\hat{R}_i^t/\Lambda_d)^2}$. If $\bar{R}_i^t\le\alpha\Lambda_d$ for some $\alpha\in[0,1)$, then
\begin{equation}
 \max\{\hat{R}_i^t,0\}\le\frac{\alpha}{\sqrt{1-\alpha^2}}\,\beta\Lambda_d.
 \label{eq:normalized-to-soft}
\end{equation}
\end{restatable}

\noindent\Cref{lem:regret-comparison} and \cref{lem:normalized-to-soft} together show that it remains to bound the normalized soft regret $\bar{R}_i^T$ by a fixed fraction of $\Lambda_d$. The following lemma gives a clear route to this bound via an RVU-style inequality~\citep{syrgkanis_rvu_2015} for normalized soft regret $\bar{R}_i^T$. %

\Needspace{8\baselineskip}
\begin{lemma}[RVU inequality for normalized soft regret]\label{lem:rvu}
For player $i$, fix arbitrary strategies $\{x_i^t\}_{t\ge1}$ and an arbitrary loss sequence $\{\ell_i^t\}_{t\ge1}\subseteq[0,1]^{d_i}$. Suppose $\beta_i^1=\beta\ge4$ and the step sizes follow $\beta_i^{t+1}=\sqrt{\beta^2+(\hat{R}_i^t/\Lambda_d)^2}$ for every $t\ge1$. Then, for every $T\ge1$,
\begin{equation}
 \bar{R}_i^T
 \le\frac{3\beta^2}{2}\sum_{t=1}^T\frac{\KL(x_i^t\|p_i^t)}{(\beta_i^t)^2}
 -\frac{\beta^2}{2}\sum_{t=1}^T\frac{\KL(x_i^t\|q_i^t)}{(\beta_i^t)^2}
 \label{eq:rvu}
\end{equation}
where $p_i^t\coloneqq\softmax(-L_i^t/\beta_i^t)$ and $q_i^t\coloneqq\softmax(-L_i^{t-1}/\beta_i^t)$.
\end{lemma}

We emphasize that this bound holds for any sequence of $x_i^t$ and is mainly a property of the normalized soft regret.
If we plug in the strategy
$x_i^{t}$ from CAOH, that is, $\softmax\left(-\frac{L_i^{t-1}+\ell_i^{t-1}}{\beta_i^{t}}\right)$,
we see that $\KL(x_i^t\|p_i^t)$ is a way to measure the difference between $\ell_i^{t-1}$ and $\ell_i^{t}$, or in other words, the error of predicting the next loss vector $\ell_i^{t}$ using the most recent loss $\ell_i^{t-1}$.
On the other hand, $q_i^t$ is an auxiliary strategy that removes the optimism from the strategy $x_t$, and $\KL(x_i^t\|q_i^t)$ measures the discrepancy between them.
More formally, in \Cref{lem:kl-bounds}, we show that $\KL(x_i^t\|p_i^t)\asymp\Var_{x_i^t}(\ell_i^t-\ell_i^{t-1})/(\beta_i^t)^2$ and $\KL(x_i^t\|q_i^t)\asymp\Var_{x_i^t}(\ell_i^{t-1})/(\beta_i^t)^2$ where $\Var_u(z)=\sum_a u_a(z_a-\ip uz)^2$ is the variance of a random coordinate (drawn from $u$) of $z$.
Therefore, with the particular form of $x_t$ from CAOH,
\Cref{lem:rvu} becomes a weight version of the variance-refined RVU inequality of~\citet[Lemma~4.1]{daskalakis_fishelson_golowich_2021}, with the important distinction that it is for the normalized soft regret $\bar{R}_i^T$ instead of the standard regret $R_i^T$.

The next lemma bounds the prediction error term $\sum_{t=1}^T\KL(x_i^t\|p_i^t)$ in terms of $\KL(x_i^t\|q_i^t)$ under self-play.

\begin{lemma}[Prediction error bound under self-play]\label{lem:prediction}
If every player uses \algorithmref{alg:caoh} with $\beta\ge4$, then, for every horizon $T\ge0$, we have
\begin{equation}
 \max_i\sum_{t=1}^T\KL(x_i^t\|p_i^t)
 \le \frac{17}{4(\beta^2-13)}
     +\frac{960n\beta^4}{(\beta^2-13)^2}
          \max_i\sum_{t=1}^T\frac{\KL(x_i^t\|q_i^t)}{(\beta_i^t)^2}.
 \label{eq:prediction}
\end{equation}
\end{lemma}
A direct combination of \cref{lem:rvu} and \cref{lem:prediction}, together with the important fact that normalized soft regret cannot be too negative,  proves our main result \cref{thm:main}, as shown below.

\begin{proof}\textbf{(\Cref{thm:main})}
Let $i$ maximize $\sum_{t=1}^T\KL(x_i^t\|p_i^t)$ over players. For any player $j$, by definition we have $(\bar{R}_j^T/\Lambda_d)^2=1-(\beta/\beta_j^{T+1})^2<1$, so $\bar{R}_j^T>-\Lambda_d$. Applying \cref{lem:rvu} to player $j$ gives
\begin{equation}
 \beta^2\sum_{t=1}^T\frac{\KL(x_j^t\|q_j^t)}{(\beta_j^t)^2}
 \le 2\Lambda_d+3\beta^2\sum_{t=1}^T
        \frac{\KL(x_j^t\|p_j^t)}{(\beta_j^t)^2}
 \le 2\Lambda_d+3\sum_{t=1}^T\KL(x_i^t\|p_i^t),
 \label{eq:weighted-kl-bound}
\end{equation}
where the last inequality uses $\beta_j^t\ge\beta$ and the choice of $i$.
For $\beta=128\sqrt n$, we have $\beta^2-13\ge(99/100)\beta^2$. Thus the constant term and the coefficient of the maximum in \eqref{eq:prediction} are at most $6/\beta^2$ and $1024n$, respectively. Combining \eqref{eq:prediction} with \eqref{eq:weighted-kl-bound} and using $1024n/\beta^2=1/16$ gives
\begin{align*}
 \sum_{t=1}^T\KL(x_i^t\|p_i^t)
 &\le \frac6{\beta^2}+1024n\max_j\sum_{t=1}^T
             \frac{\KL(x_j^t\|q_j^t)}{(\beta_j^t)^2}\\
 &\le\frac6{\beta^2}+\frac{\Lambda_d}{8}
       +\frac3{16}\sum_{t=1}^T\KL(x_i^t\|p_i^t).
\end{align*}
Rearranging and using $\Lambda_d\ge16$ yields
\begin{equation}
 \sum_{t=1}^T\KL(x_i^t\|p_i^t)
 \le\frac{2\Lambda_d}{13}+\frac{96}{13\beta^2}
 \le\frac{\Lambda_d}{4}.
 \label{eq:closure}
\end{equation}
By \cref{lem:rvu} and $\beta_j^t\ge\beta$, $\bar{R}_j^T\le\frac32\sum_{t=1}^T\KL(x_i^t\|p_i^t)\le3\Lambda_d/8\le\Lambda_d/2$ for every player. Applying \cref{lem:normalized-to-soft} with $\alpha=1/2$ and using $1/\sqrt3\le2/3$ gives $\max\{\hat{R}_j^T, 0\} \leq \frac{2}{3}\beta\Lambda_d$.
Further using \cref{lem:regret-comparison} gives $\Reg_j^T\le7\beta\Lambda_d/8<\beta\Lambda_d$, which finishes the proof (since $\Lambda_d\le40\log d$ for any $d\geq 2$).
\end{proof}

\noindent We next sketch the proofs for \cref{lem:rvu} and \cref{lem:prediction} in \Cref{sec:normalized} and \Cref{sec:prediction}, respectively.

\subsection{RVU inequality for normalized soft regret}\label{sec:normalized}
To prove the RVU inequality in \cref{lem:rvu}, we start from a one-step inequality for normalized soft regret difference (omitted proofs for this subsection are deferred to Appendix~\ref{app:rvu-proof}).
\begin{restatable}[One-step RVU for normalized soft regret difference]{lemma}{stepnormalized}\label{lem:normalized-step}
For player $i$, fix arbitrary strategies $\{x_i^t\}_{t\ge1}$ and an arbitrary loss sequence $\{\ell_i^t\}_{t\ge1}\subseteq[0,1]^{d_i}$. Suppose $\beta_i^1=\beta\ge4$ and the step sizes follow $\beta_i^{t+1}=\sqrt{\beta^2+(\hat{R}_i^t/\Lambda_d)^2}$ for every $t\ge1$. Then, for every $t\ge1$,
\begin{equation}
 \bar{R}_i^t-\bar{R}_i^{t-1}
 \le\frac{\beta^2}{(\beta_i^t)^2}\left(
 \frac87\KL(x_i^t\|p_i^t)-\frac78\KL(x_i^t\|q_i^t)
 +\frac{8\log d_i}{7}\frac{|\beta_i^t-\beta_i^{t-1}|}{\beta_i^t}\right).
 \label{eq:normalized-step}
\end{equation}
\end{restatable}
\Cref{lem:normalized-step} suggests summing the one-step inequality to obtain \cref{lem:rvu}. However, the last term in \eqref{eq:normalized-step}, which accounts for changes in inverse step size, can offset the negative KL term and loosen the upper bound. The next lemma controls these changes in terms of $\KL(x_i^t\|p_i^t)$ and $\KL(x_i^t\|q_i^t)$, and rules out complete cancellation of the negative KL term after summation. We note that only \eqref{eq:regularization-rate-changes} is needed for \Cref{lem:rvu} and \eqref{eq:trajectory-rates} will be needed for results in the next subsection.
\begin{restatable}[Bounds on inverse step sizes]{lemma}{weightedrates}\label{lem:weighted-rates}
For player $i$, fix arbitrary strategies $\{x_i^t\}_{t\ge1}$ and an arbitrary loss sequence $\{\ell_i^t\}_{t\ge1}\subseteq[0,1]^{d_i}$. Suppose $\beta_i^1=\beta\ge4$ and the step sizes follow $\beta_i^{t+1}=\sqrt{\beta^2+(\hat{R}_i^t/\Lambda_d)^2}$ for every $t\ge1$. Then, for every $T\ge1$,
\begin{align}
 (\log d_i)\sum_{t=1}^{T}\frac{|\beta_i^t-\beta_i^{t-1}|}{(\beta_i^t)^3}
 &\le\frac18\sum_{t=1}^{T}
 \frac{\KL(x_i^t\|p_i^t)+\KL(x_i^t\|q_i^t)}{(\beta_i^t)^2},
 \label{eq:regularization-rate-changes}\\
 \Lambda_d^2\sum_{t=1}^{T-1}\left(\log\frac{\beta_i^{t+1}}{\beta_i^t}\right)^2
 &\le2\sum_{t=1}^{T}
 \frac{\KL(x_i^t\|p_i^t)+\KL(x_i^t\|q_i^t)}{(\beta_i^t)^2}.
 \label{eq:trajectory-rates}
\end{align}
\end{restatable}
\begin{proof}\textbf{(\Cref{lem:rvu})}
Summing \eqref{eq:normalized-step} over $t=1,\ldots,T$ and using $\bar{R}_i^0=0$ and \eqref{eq:regularization-rate-changes} gives
\begin{align*}
 \bar{R}_i^T
 &\le\beta^2\sum_{t=1}^T\frac{1}{(\beta_i^t)^2}
 \left(\frac87\KL(x_i^t\|p_i^t)-\frac78\KL(x_i^t\|q_i^t)
 +\frac{\KL(x_i^t\|p_i^t)+\KL(x_i^t\|q_i^t)}7\right)\\*
 &\le\frac{3\beta^2}{2}\sum_{t=1}^T\frac{\KL(x_i^t\|p_i^t)}{(\beta_i^t)^2}
 -\frac{\beta^2}{2}\sum_{t=1}^T\frac{\KL(x_i^t\|q_i^t)}{(\beta_i^t)^2},
 \tag{$\frac87+\frac17\le\frac32$ and $\frac78-\frac17\ge\frac12$}
\end{align*}
which proves \Cref{lem:rvu}.
\end{proof}

\subsection{Prediction error bound under self-play}\label{sec:prediction}
Now it remains to prove the prediction error bound in \cref{lem:prediction}. 
To this end, we define the first-order difference of the loss vector sequence as $\Delta\ell_i^t=\ell_i^t-\ell_i^{t-1}$ and the second-order difference as $\Delta^2\ell_i^t=\ell_i^{t+1}-2\ell_i^t+\ell_i^{t-1}=\Delta\ell_i^{t+1}-\Delta\ell_i^t$ for $t\ge1$. For each player $i$ and horizon $T$, set
\begin{equation}
 P_i(T)\coloneqq\sum_{t=1}^T\KL(x_i^t\|p_i^t),\quad
 Q_i(T)\coloneqq\sum_{t=1}^T\frac{\KL(x_i^t\|q_i^t)}{(\beta_i^t)^2},\quad
 D_i(T)\coloneqq\left(\sum_{t=2}^{T-1}\|\Delta^2\ell_i^t\|_\infty^2\right)^{1/2}.
 \label{eq:prediction-sums}
\end{equation}
The next lemma bounds $P_i(T)$ in terms of $Q_i(T)$ and $D_i(T)$ for arbitrary loss sequences (omitted proofs for this subsection are deferred to Appendix~\ref{app:prediction-proof}).
Notably, while several earlier results rely crucially on analyzing high-order differences of the loss sequence~\citep{daskalakis_fishelson_golowich_2021, liu_farina_ozdaglar_2026, abbadi_laraki_mertikopoulos_2026, ha_horizon_hedge_2026, ha_log_hedge_2026},
here we only need to analyze the second-order difference.

\begin{restatable}[Prediction error bound]{lemma}{reductionlemma}\label{lem:reduction}
Fix player $i$ using \algorithmref{alg:caoh} with $\beta\ge4$ for an arbitrary loss sequence $\{\ell_i^t\}_{t\ge1}\subseteq[0,1]^{d_i}$. Then for every integer $T\ge1$,
\begin{equation}
 \left(1-\frac{13}{\beta^2}\right)P_i(T)
 \le \frac{17}{8\beta^2}+51Q_i(T)+4\sqrt{Q_i(T)}\,D_i(T).
 \label{eq:reduction}
\end{equation}
\end{restatable}

Under self-play, the second-order loss differences $\Delta^2\ell_i^t$ satisfy the following bound.

\begin{restatable}[Smoothness under self-play]{lemma}{losslemma}\label{lem:loss-differences}
Fix player $i$. If its opponents use \algorithmref{alg:caoh} with $\beta\ge4$ and $\ell_i^t=\ell_i(x_{-i}^t)$ for every round $t$, then for every integer $T\ge1$,
\begin{equation}
 D_i(T)\le
 \left(3+\sqrt{20}+\frac1{8\beta}\right)\sqrt{\sum_{j\ne i}P_j(T)}
 +4\sum_{j\ne i}\sqrt{Q_j(T)}.
 \label{eq:accumulated-loss-differences}
\end{equation}
\end{restatable}

Now, we are ready to prove the prediction error bound in \cref{lem:prediction}.

\begin{proof}\textbf{(\Cref{lem:prediction})}
Fix any $T$.
Let $i$ maximize $\sum_{t=1}^T\KL(x_i^t\|p_i^t)$ over players, and let $j$ maximize $\sum_{t=1}^T\KL(x_j^t\|q_j^t)/(\beta_j^t)^2$. Write $\kappa_\beta=3+\sqrt{20}+1/(8\beta)$. Substituting \eqref{eq:accumulated-loss-differences} into \eqref{eq:reduction} for player $i$ and bounding each opponent's sum by the corresponding maximum gives
\begin{align*}
 \left(1-\frac{13}{\beta^2}\right)\sum_{t=1}^T\KL(x_i^t\|p_i^t)
 &\le\frac{17}{8\beta^2}+[51+16(n-1)]
              \sum_{t=1}^T\frac{\KL(x_j^t\|q_j^t)}{(\beta_j^t)^2}\\
 &\quad +4\kappa_\beta
 \left((n-1)\sum_{t=1}^T\KL(x_i^t\|p_i^t)
             \sum_{t=1}^T\frac{\KL(x_j^t\|q_j^t)}{(\beta_j^t)^2}\right)^{1/2}.
\end{align*}
Apply AM-GM inequality $2\sqrt{ab}\le a+b$, 
we can bound the last term by
\[
 \frac{\beta^2-13}{2\beta^2}\sum_{t=1}^T\KL(x_i^t\|p_i^t)
 +\frac{8\kappa_\beta^2\beta^2(n-1)}{\beta^2-13}
      \sum_{t=1}^T\frac{\KL(x_j^t\|q_j^t)}{(\beta_j^t)^2}.
\]
Subtracting $(\beta^2-13)/(2\beta^2)$ times $\sum_{t=1}^T\KL(x_i^t\|p_i^t)$ leaves the same positive coefficient on the left, since $\beta\ge4$. Dividing by that coefficient and using $\beta^2-13\le\beta^2$ gives
\begin{align*}
 \sum_{t=1}^T\KL(x_i^t\|p_i^t)
 &\le\frac{17}{4(\beta^2-13)}
   +\frac{\beta^4[102+(32+16\kappa_\beta^2)(n-1)]}{(\beta^2-13)^2}
      \sum_{t=1}^T\frac{\KL(x_j^t\|q_j^t)}{(\beta_j^t)^2}\\
 &\le\frac{17}{4(\beta^2-13)}+\frac{960n\beta^4}{(\beta^2-13)^2}
      \sum_{t=1}^T\frac{\KL(x_j^t\|q_j^t)}{(\beta_j^t)^2},
\end{align*}
where the last line uses $\kappa_\beta<38/5$ for $\beta\ge4$.
This proves \Cref{lem:prediction}.
\end{proof}

\section{Conclusion}
We established a constant $O(\sqrt n\log d)$ individual external regret bound for Cautiously Adaptive Optimistic Hedge in general-sum games. Our adaptive step sizes keep normalized soft regret bounded below, which, together with the smoothness of self-play-induced loss sequences, allows us to bound the accumulated prediction error by a constant. As a result, CAOH avoids the $2^{\Omega(nd)}$ regret lower bound that constant step sizes can suffer~\citep{dagan_eisen_erez_koren_2026}, while retaining the original Optimistic Hedge update. We leave the optimality of CAOH as an intriguing open question.

\section*{AI use statement}
Motivated by the horizon-independent regret bound of \citet{ha_horizon_hedge_2026} and an exponential lower bound we found independently of \citet{dagan_eisen_erez_koren_2026} in a separate discussion with ChatGPT~6~Pro, we asked the model whether Optimistic Hedge could attain an $O(\operatorname{poly}(n)\log d)$ constant regret bound under self-play using simple closed-form adaptive step sizes.
It proposed several candidates, including a highly unintuitive rule that used the current strategy $x_i^t$ as a comparator in place of the best entropy-regularized strategy.
Inspired by cautious optimism~\citep{soleymani_cautious_2025,soleymani_cautious_ec_2025} and recent constant-regret results based on lifted regularization~\citep{liu_farina_ozdaglar_2026,abbadi_laraki_mertikopoulos_2026}, we restricted the search to updates that (1) reduce the step sizes under excessive performance, and (2) use an intuitive, efficiently computable statistic as a benchmark with the goal of an efficient implementation in extensive-form games.
Further discussions with ChatGPT~6~Pro under these restrictions led to CAOH.
The model generated the initial proof of \cref{thm:main} and a proof of \cref{thm:ordinary-linear}, which is omitted here, and we refactored the former into a chain of interpretable lemmas.
We designed the experiment in Figure~\ref{fig:dimension-sweeps}, which ChatGPT~6.1~Sol implemented and conducted.

\begingroup
\setlength{\bibsep}{4pt plus 2pt minus 1pt}
\hypersetup{urlcolor=black}
\bibliography{references}
\endgroup

\newpage
\appendix
\section{Basic Lemmas}\label{app:preliminary-lemmas}
This section provides useful helper lemmas for our proofs.
Throughout, $\sqrt{u}$ for a vector $u$ represents taking square root coordinate-wise.

\begin{lemma}[Entropic minimizer]\label{lem:entropy-minimum}
\setlength{\emergencystretch}{2em}
For $k\ge1$, $b>0$, and $L\in\R^k$, define
$\Phi_b(L)=-b\log(k^{-1}\sum_{a\in[k]} e^{-L_a/b})$ and $z=\softmax(-L/b)$. Then, for every $u\in\Delta_k$,
\begin{equation}
 \ip uL+b\KL(u\|\one/k)=\Phi_b(L)+b\KL(u\|z).
 \label{eq:regularized-loss-identity}
\end{equation}
Also, $z$ uniquely minimizes the objective in \eqref{eq:regularized-loss-identity} as a function of $u$, and
$\partial_b\Phi_b(L)=\KL(z\|\one/k)\in[0,\log k]$ holds.
\end{lemma}
\begin{proof}
Substitute $\log z_a=-L_a/b-\log\sum_{c\in[k]} e^{-L_c/b}$ into the KL divergence proves \eqref{eq:regularized-loss-identity}. 
The optimality of $z$ is immediate by considering the right-hand size of \eqref{eq:regularized-loss-identity} and using the nonnegativity of KL.
Finally, the formula for the derivative is by direct calculation.
\end{proof}

\Needspace{7\baselineskip}
\noindent For $z\in\R^k$ and $u\in\Delta_k$, write $\operatorname{osc}(z)=\max_a z_a-\min_a z_a$ and $\Var_u(z)=\sum_a u_a(z_a-\ip uz)^2$.
\begin{lemma}[One-step KL divergence as variance]\label{lem:reweighting-kl}
Let $u\in\Delta_k$ be positive in all coordinates, $z\in\R^k$, and $v=\softmax(\log u+z)$. Then
$\KL(u\|v)\le\operatorname{osc}(z)^2/8$. Moreover, if $\operatorname{osc}(z)\le1/2$, then
\begin{equation}
 \frac14\Var_u(z)\le\KL(u\|v)\le\Var_u(z).
 \label{eq:generic-kl-variance}
\end{equation}
\end{lemma}
\begin{proof}
Define $F(s)=\log\sum_{a\in[k]}u_a e^{s z_a}$ and $u(s)=\softmax(\log u+s z)$ for $0\le s\le1$.
The definition of softmax gives $u(s)_a=u_a e^{s z_a-F(s)}$, with $u(0)=u$ and $u(1)=v$.
Differentiation gives $F'(s)=\ip{u(s)}{z}$ and $F''(s)=\sum_a u(s)_a z_a^2-F'(s)^2=\Var_{u(s)}(z)$. Hence
\begin{align}
 \KL(u\|v)
 &=\sum_a u_a\log\frac{u_a}{v_a}
 =\sum_a u_a\bigl(F(1)-z_a\bigr)\nonumber\\
 &=F(1)-F(0)-F'(0)
 \tag{$F(0)=0$, $F'(0)=\ip{u}{z}$}\\
 &=\int_0^1(1-s)F''(s)\,ds
 \tag{Taylor's theorem}\\
 &=\int_0^1(1-s)\Var_{u(s)}(z)\,ds.
 \label{eq:kl-integral}
\end{align}
The variance bound $\Var_{u(s)}(z)\le\operatorname{osc}(z)^2/4$ then proves the first claim.
Now suppose $\operatorname{osc}(z)\le1/2$. The denominator of $u(s)_a/u_a=e^{s z_a}/(\sum_b u_b e^{s z_b})$ is a weighted average of the $e^{s z_b}$, so the ratio lies between $e^{-s\operatorname{osc}(z)}$ and $e^{s\operatorname{osc}(z)}$.
Thus $u_a/2\le u(s)_a\le2u_a$, since $s\operatorname{osc}(z)\le1/2$ and $e^{1/2}<2$.
Multiplying these bounds by $(z_a-c)^2$ and summing over $a$ gives, for every $c\in\R$,
\[
 \frac12\sum_a u_a(z_a-c)^2
 \le\sum_a u(s)_a(z_a-c)^2
 \le2\sum_a u_a(z_a-c)^2.
\]
Since $\Var_w(z)=\min_{c\in\R}\sum_a w_a(z_a-c)^2$ for any $w\in\Delta_k$, taking the minimum over $c$ in each term yields $\Var_u(z)/2\le\Var_{u(s)}(z)\le2\Var_u(z)$.
Substituting these inequalities into \eqref{eq:kl-integral} gives \eqref{eq:generic-kl-variance}.
\end{proof}

\Needspace{6\baselineskip}
\begin{lemma}[Hellinger--KL inequality]\label{lem:probability-comparisons}
For probability vectors $u,v\in\Delta_k$, with $-\log0=+\infty$ and the extended-value convention for KL when coordinates vanish, we have
\begin{equation}
 \norm{\sqrt u-\sqrt v}_2^2
 =2\left(1-\sum_{a\in[k]}\sqrt{u_av_a}\right)
 \le-2\log\sum_{a\in[k]}\sqrt{u_av_a}\le\KL(u\|v).
 \label{eq:kl-distance}
\end{equation}
\end{lemma}
\begin{proof}
Expanding the square and using $\sum_a u_a=\sum_a v_a=1$ gives
\[
 \norm{\sqrt u-\sqrt v}_2^2
 =\sum_{a\in[k]}(u_a+v_a-2\sqrt{u_av_a})
 =2\left(1-\sum_{a\in[k]}\sqrt{u_av_a}\right).
\]
The first inequality in \eqref{eq:kl-distance} follows from $-\log z\ge1-z$ for $z\ge0$.
If $v_a=0$ for some $a$ with $u_a>0$, then $\KL(u\|v)=+\infty$ and the last inequality holds. Otherwise, $\sqrt{u_av_a}=u_a\sqrt{v_a/u_a}$ for $u_a>0$, whereas the terms with $u_a=0$ contribute zero. On the support of $u$, the square roots are positive and the weights $u_a$ sum to $1$. Jensen's inequality for $-\log$ gives
\[
 -2\log\sum_{a\in[k]}\sqrt{u_av_a}
 \le-2\sum_{a:u_a>0}u_a\log\sqrt{v_a/u_a}
 =\sum_{a:u_a>0}u_a\log(u_a/v_a)=\KL(u\|v).
\]
\end{proof}

\begin{lemma}[Hellinger bound for product distributions]\label{lem:product-comparisons}
For probability vectors $u_j,v_j\in\Delta_{d_j}$, product distributions $P=\prod_j u_j$ and $Q=\prod_j v_j$, and $h:\prod_j[d_j]\to[0,1]$,
\begin{equation}
 |\E_Ph-\E_Qh|^2
 \le\tfrac14\|P-Q\|_1^2
 \le\norm{\sqrt P-\sqrt Q}_2^2
 \le\sum_j\norm{\sqrt{u_j}-\sqrt{v_j}}_2^2.
 \label{eq:product}
\end{equation}
\end{lemma}
\begin{proof}
Boundedness of $h$ gives the first inequality, and Cauchy--Schwarz with $\|\sqrt P+\sqrt Q\|_2\le2$ gives the second.
For the last inequality, let $a_j=\ip{\sqrt{u_j}}{\sqrt{v_j}}$. Then, each $a_j$ lies in $[0,1]$ and $\ip{\sqrt P}{\sqrt Q}=\prod_j a_j$ holds.
Applying the identity in \cref{lem:probability-comparisons} to the joint distributions and to each pair of marginals gives
\begin{align*}
 \norm{\sqrt P-\sqrt Q}_2^2
 &=2\left(1-\prod_j a_j\right)
 =2\sum_j(1-a_j)\prod_{r<j}a_r\\*
 &\le2\sum_j(1-a_j)
 =\sum_j\norm{\sqrt{u_j}-\sqrt{v_j}}_2^2.
\end{align*}
Here, the sum in the first line telescopes because $(1-a_j)\prod_{r<j}a_r=\prod_{r<j}a_r-\prod_{r\le j}a_r$, with the empty product equal to $1$, and the inequality uses $\prod_{r<j}a_r\le1$.
\end{proof}

\Needspace{7\baselineskip}
\begin{lemma}[Logit rescaling]\label{lem:rescaling}
\setlength{\emergencystretch}{2em}
For a probability vector $u\in\Delta_k$ and $a>0$, we have
\begin{equation}
 \norm{\sqrt{\softmax(a\log u)}-\sqrt u}_2^2
 \le\frac{(1+\log k)^2}{2}(\log a)^2.
 \label{eq:rescaling-distance}
\end{equation}
\end{lemma}
\begin{proof}
To bound the distance in \eqref{eq:rescaling-distance}, let $u(s)=\softmax(a^s\log u)$ for $0\le s\le1$.
Set $Z(s)=\sum_{c:u_c>0}u_c^{a^s}$ and define $u(s)_b=u_b^{a^s}/Z(s)$ for $u_b>0$ and $u(s)_b=0$ for $u_b=0$.
Then $u(0)=u$ and $u(1)=\softmax(a\log u)$. By the fundamental theorem of calculus and the triangle inequality,
\[
 \norm{\sqrt{u(1)}-\sqrt{u(0)}}_2
 \le\int_0^1\left\|\frac d{ds}\sqrt{u(s)}\right\|_2\,ds.
\]
Thus it suffices to bound the integrand by $(1+\log k)|\log a|/\sqrt{2}$ for every $s$.
For $u_b>0$, differentiating $\log u(s)_b=a^s\log u_b-\log Z(s)$ and using $\sum_c u(s)_c=1$ gives
\begin{align*}
 \frac{d}{ds}\log u(s)_b
 &=(\log a)\left(a^s\log u_b-\sum_{c:u_c>0}u(s)_c a^s\log u_c\right)\\*
 &=(\log a)\left(\log u(s)_b-\sum_{c:u_c>0}u(s)_c\log u(s)_c
       +\left(1-\sum_c u(s)_c\right)\log Z(s)\right)\\*
 &=(\log a)\left(\log u(s)_b-\sum_{c:u_c>0}u(s)_c\log u(s)_c\right).
\end{align*}
Therefore, it follows from the chain rule that
\begin{align*}
 \left\|\frac d{ds}\sqrt{u(s)}\right\|_2
 &=\sqrt{\sum_{b:u_b>0}\left(\frac{d}{ds}\sqrt{u(s)_b}\right)^2}
 =\frac12\sqrt{\sum_{b:u_b>0}u(s)_b\left(\frac{d}{ds}\log u(s)_b\right)^2}\\*
 &=\frac{|\log a|}{2}\sqrt{\sum_{b:u_b>0}u(s)_b
        \left(\log u(s)_b-\sum_{c:u_c>0}u(s)_c\log u(s)_c\right)^2}\\*
 &=\frac{|\log a|}{2}\sqrt{\Var_{u(s)}(\log u(s))}.
\end{align*}

It remains to bound $\Var_{u(s)}(\log u(s))$ by $2(1+\log k)^2$ for every $s$. Fix $s$, let $B$ have distribution $u(s)$, and set $Y=-\log u(s)_B$, which is finite and nonnegative almost surely.
For $t\ge\log k$, $Y\ge t$ means $u(s)_B\le e^{-t}$, so summing over at most $k$ such coordinates gives $\Pr(Y\ge t)\le ke^{-t}$.
Splitting the tail formula $\E[Y^2]=\int_0^\infty 2t\Pr(Y\ge t)\,dt$ at $\log k$ and using $\Pr(Y\ge t)\le1$ below that point gives
\begin{align}
 \Var_{u(s)}(\log u(s))
 &\le \E[Y^2]\le \int_0^{\log k}2t\,dt+\int_{\log k}^{\infty}2tke^{-t}\,dt\notag\\*
 &=(\log k)^2+2\log k+2\le2(1+\log k)^2.
 \label{eq:entropy-moment}
\end{align}
This gives the required bound on the integrand. Integrating over $[0,1]$ and squaring proves \eqref{eq:rescaling-distance}.
\end{proof}

\Needspace{5\baselineskip}
\section{Soft Regret}\label{app:preliminary-estimates}
In this section, we prove \Cref{lem:regret-comparison} and \Cref{lem:normalized-to-soft} after introducing some helper lemmas.
\phantomsection
\begin{lemma}[Lipschitz bound for the inverse step size update]\label{lem:inverse-step-lipschitz}
For $\beta>0$ and $\Lambda_d>0$, the map $b(g)=\sqrt{\beta^2+(g/\Lambda_d)^2}$ is $1/\Lambda_d$-Lipschitz on $\R$.
\end{lemma}
\begin{proof}
For $g,h\in\R$, writing $b(g)=\norm{(\beta,g/\Lambda_d)}_2$, the reverse triangle inequality gives
\[
 |b(g)-b(h)|\le\norm{(0,(g-h)/\Lambda_d)}_2=\frac{|g-h|}{\Lambda_d}.
\]
\end{proof}

\phantomsection
\begin{lemma}[Stability of inverse step sizes]\label{lem:rates}
\setlength{\emergencystretch}{2em}
For player $i$, fix arbitrary strategies $\{x_i^t\}_{t\ge1}$ and an arbitrary loss sequence $\{\ell_i^t\}_{t\ge1}\subseteq[0,1]^{d_i}$. Suppose $\beta_i^1=\beta\ge4$ and the step sizes follow $\beta_i^{t+1}=\sqrt{\beta^2+(\hat{R}_i^t/\Lambda_d)^2}$ for every $t\ge1$. Then, for every $t\ge1$,
\begin{equation}
 |\beta_i^{t+1}-\beta_i^t|
 \le\frac{|\hat{R}_i^t-\hat{R}_i^{t-1}|}{\Lambda_d}\le\frac1{16},\qquad
 \max\left\{\frac{\beta_i^{t+1}}{\beta_i^t},\frac{\beta_i^t}{\beta_i^{t+1}}\right\}\le\frac{21}{20}.
 \label{eq:generic-rates}
\end{equation}
\end{lemma}
\begin{proof}
Fix $i$ and let $\delta_t=|\beta_i^t-\beta_i^{t-1}|$, so $\delta_1=0$.
At a fixed inverse step size, adding $\ell_i^t$ increases the regularized minimum in \eqref{eq:soft-regret} by a number in $[0,1]$, since every comparator incurs a loss in $[0,1]$.
At fixed cumulative loss, \cref{lem:entropy-minimum} bounds the change in the minimum by $(\log d_i)\delta_t$.
The played loss also lies in $[0,1]$, so the definition of soft regret and \cref{lem:inverse-step-lipschitz} give
\[
 \Lambda_d\delta_{t+1}
 \le|\hat{R}_i^t-\hat{R}_i^{t-1}|
 \le1+(\log d_i)\delta_t.
\]
If $\delta_t\le1$, dividing by $\Lambda_d$ bounds both $\delta_{t+1}$ and $|\hat{R}_i^t-\hat{R}_i^{t-1}|/\Lambda_d$ by $(1+\log d_i)/\Lambda_d\le1/16$.
Induction from $\delta_1=0$ therefore proves the first chain in \eqref{eq:generic-rates}.
Since all inverse step sizes are at least $\beta\ge4$, it also gives
\[
 \max\left\{\frac{\beta_i^{t+1}}{\beta_i^t},\frac{\beta_i^t}{\beta_i^{t+1}}\right\}
 =1+\frac{\delta_{t+1}}{\min\{\beta_i^t,\beta_i^{t+1}\}}
 \le1+\frac{1}{16\beta}\le\frac{65}{64}<\frac{21}{20}.
\]

\end{proof}

\Needspace{4\baselineskip}
We are now ready to prove the two lemmas from \Cref{sec:reduction}.

\phantomsection\label{app:comparison-proof}
\regretcomparison*
\begin{proof}
The minimum in \eqref{eq:soft-regret} is at least $\min_a L_{i,a}^t$ by nonnegativity of KL, and at most $\min_a L_{i,a}^t+\beta_i^t\log d_i$ by testing a best action. Subtracting these bounds from the cumulative played loss gives \eqref{eq:regret-comparator}.

For \eqref{eq:regret-target}, the ratio bound in \cref{lem:rates} gives $\beta_i^t\le(21/20)\beta_i^{t+1}\le2\beta_i^{t+1}$.
Applying \cref{lem:inverse-step-lipschitz} at $\hat{R}_i^t$ and $0$ gives $\beta_i^{t+1}\le\beta+|\hat{R}_i^t|/\Lambda_d$. Using $2\log d_i\le\Lambda_d/8$ in \eqref{eq:regret-comparator}, we obtain
\begin{equation*}
 \Reg_i^t
 \le \hat{R}_i^t+\frac{\Lambda_d}{8}\beta_i^{t+1}
 \le\frac{\beta\Lambda_d}{8}+\hat{R}_i^t+\frac18|\hat{R}_i^t|
 \le\frac{\beta\Lambda_d}{8}+\frac98\max\{\hat{R}_i^t,0\}. \tag*{\BlackBox}
\end{equation*}
\renewcommand{\jmlrQED}{\par}
\end{proof}

\Needspace{5\baselineskip}
\phantomsection\label{app:normalized-to-soft}
\normalizedsoft*
\begin{proof}
If $\hat{R}_i^t\le0$, the bound is immediate. Otherwise, the definition of $\bar{R}_i^t$ and the assumed formula for $\beta_i^{t+1}$ give
$(\hat{R}_i^t)^2\le\alpha^2\Lambda_d^2(\beta_i^{t+1})^2
 =\alpha^2\bigl(\beta^2\Lambda_d^2+(\hat{R}_i^t)^2\bigr)
$, which implies $\hat{R}_i^t\le\frac{\alpha\beta\Lambda_d}{\sqrt{1-\alpha^2}}$.
\end{proof}

\Needspace{5\baselineskip}
\section{Proof of the RVU Inequality}\label{app:rvu-proof}

In this section, we prove the two main lemmas from \Cref{sec:normalized}.
Recall the definition $p_i^t\coloneqq\softmax(-L_i^t/\beta_i^t)$ and $q_i^t\coloneqq\softmax(-L_i^{t-1}/\beta_i^t)$ from \Cref{lem:rvu}.

\begin{lemma}[One-step soft regret decomposition]\label{lem:soft-regret-decomposition}
For player $i$, fix arbitrary strategies $\{x_i^t\}_{t\ge1}$, an arbitrary loss sequence $\{\ell_i^t\}_{t\ge1}\subseteq[0,1]^{d_i}$, and positive inverse step sizes $\{\beta_i^t\}_{t\ge1}$. Then there exists a sequence $\{a_i^t\}_{t\ge1}\subseteq[0,\log d_i]$ such that the following holds for every $t\ge1$:
\begin{equation}
 \hat{R}_i^t-\hat{R}_i^{t-1}
 =\beta_i^t\bigl[\KL(x_i^t\|p_i^t)-\KL(x_i^t\|q_i^t)\bigr]
 -a_i^t(\beta_i^t-\beta_i^{t-1}).
 \label{eq:soft-regret-increment}
\end{equation}
\end{lemma}
\begin{proof}
We separate the change in cumulative loss from the change in inverse step size by evaluating both $L_i^t$ and $L_i^{t-1}$ at $\beta_i^t$.
By \cref{lem:entropy-minimum}, the minimum in \eqref{eq:soft-regret} is $\Phi_{\beta_i^t}(L_i^t)$, where $\Phi_b(L)=-b\log\left(d_i^{-1}\sum_{a\in[d_i]}e^{-L_a/b}\right)$.
Taking the difference and using $L_i^t-L_i^{t-1}=\ell_i^t$ yields
\begin{align*}
 \hat{R}_i^t-\hat{R}_i^{t-1}
 &=\ip{x_i^t}{\ell_i^t}-\Phi_{\beta_i^t}(L_i^t)+\Phi_{\beta_i^{t-1}}(L_i^{t-1})\\*
 &=\bigl[\ip{x_i^t}{L_i^t}-\Phi_{\beta_i^t}(L_i^t)\bigr]
   -\bigl[\ip{x_i^t}{L_i^{t-1}}-\Phi_{\beta_i^t}(L_i^{t-1})\bigr]\\*
 &\quad+\Phi_{\beta_i^{t-1}}(L_i^{t-1})-\Phi_{\beta_i^t}(L_i^{t-1}).
\end{align*}
By definition, $p_i^t$ and $q_i^t$ are the entropic minimizers for $L_i^t$ and $L_i^{t-1}$ at inverse step size $\beta_i^t$. Applying \eqref{eq:regularized-loss-identity} in \cref{lem:entropy-minimum} with $u=x_i^t$ and $b=\beta_i^t$ gives
\begin{align*}
 \ip{x_i^t}{L_i^t}-\Phi_{\beta_i^t}(L_i^t)
 &=\beta_i^t\bigl[\KL(x_i^t\|p_i^t)-\KL(x_i^t\|\one/d_i)\bigr],\\*
 \ip{x_i^t}{L_i^{t-1}}-\Phi_{\beta_i^t}(L_i^{t-1})
 &=\beta_i^t\bigl[\KL(x_i^t\|q_i^t)-\KL(x_i^t\|\one/d_i)\bigr].
\end{align*}
Subtracting the second line from the first gives the KL difference in \eqref{eq:soft-regret-increment}.

For the remaining difference, keep $L_i^{t-1}$ fixed and apply the mean value theorem to $b\mapsto\Phi_b(L_i^{t-1})$. When $\beta_i^t\neq\beta_i^{t-1}$, there is a point $b_\ast$ between them such that
\[
 \Phi_{\beta_i^{t-1}}(L_i^{t-1})-\Phi_{\beta_i^t}(L_i^{t-1})
 =-a_i^t(\beta_i^t-\beta_i^{t-1}),\qquad
 a_i^t=\left.\partial_b\Phi_b(L_i^{t-1})\right|_{b=b_\ast}.
\]
When $\beta_i^t=\beta_i^{t-1}$, take $b_\ast=\beta_i^t$, so both sides of the first equality are zero. In both cases, \cref{lem:entropy-minimum} gives $a_i^t\in[0,\log d_i]$, which completes the proof of \eqref{eq:soft-regret-increment}.
\end{proof}

\stepnormalized*
\begin{proof}
To derive \eqref{eq:normalized-step} from \cref{lem:soft-regret-decomposition}, it suffices to find a scalar $m^t$ such that
\begin{equation}
 \bar{R}_i^t-\bar{R}_i^{t-1}=m^t(\hat{R}_i^t-\hat{R}_i^{t-1}),\qquad
 \frac78\frac{\beta^2}{(\beta_i^t)^3}\le m^t\le\frac87\frac{\beta^2}{(\beta_i^t)^3}.
 \label{eq:secants}
\end{equation}
Indeed, multiplying the soft regret decomposition by $m^t$ would give
\begin{align*}
 \bar{R}_i^t-\bar{R}_i^{t-1}
 &=m^t\beta_i^t\bigl[\KL(x_i^t\|p_i^t)-\KL(x_i^t\|q_i^t)\bigr]
   -m^ta_i^t(\beta_i^t-\beta_i^{t-1})\\*
 &\le\frac{\beta^2}{(\beta_i^t)^2}\left(
 \frac87\KL(x_i^t\|p_i^t)-\frac78\KL(x_i^t\|q_i^t)
 +\frac{8\log d_i}{7}\frac{|\beta_i^t-\beta_i^{t-1}|}{\beta_i^t}\right).
\end{align*}
The positive KL term uses the upper bound on $m^t$, and the negative KL term uses the lower bound. The last term uses the upper bound and $0\le a_i^t\le\log d_i$ after taking absolute values.

It remains to prove \eqref{eq:secants}. Let $b(g)=\sqrt{\beta^2+(g/\Lambda_d)^2}$ be the map in \cref{lem:inverse-step-lipschitz} and set $f(g)=g/b(g)$. The update and initialization give $f(\hat{R}_i^t)=\bar{R}_i^t$ for $t\ge0$. Thus the mean value theorem yields the identity in \eqref{eq:secants} with $m^t=f'(g)$ for some $g$ between $\hat{R}_i^{t-1}$ and $\hat{R}_i^t$. If the endpoints coincide, take $g$ to be their common value.
Differentiating gives $f'(g)=(b(g)^2-g^2/\Lambda_d^2)/b(g)^3=\beta^2/b(g)^3$, so it remains to compare $b(g)$ with $b(\hat{R}_i^{t-1})=\beta_i^t$. By \cref{lem:inverse-step-lipschitz} and \cref{lem:rates},
\[
 \left|\frac{b(g)}{\beta_i^t}-1\right|
 \le\frac{|g-\hat{R}_i^{t-1}|}{\Lambda_d\beta_i^t}
 \le\frac{|\hat{R}_i^t-\hat{R}_i^{t-1}|}{\Lambda_d\beta_i^t}
 \le\frac{1}{16\beta_i^t}\le\frac1{64}<\frac1{32},
\]
where we used $\beta_i^t\ge4$. Hence $31/32<b(g)/\beta_i^t<33/32$, and
\[
 \frac78
 <\left(\frac{32}{33}\right)^3
 <\frac{m^t}{\beta^2/(\beta_i^t)^3}
 =\left(\frac{\beta_i^t}{b(g)}\right)^3
 <\left(\frac{32}{31}\right)^3
 <\frac87.
\]
This proves \eqref{eq:secants} and completes the proof.
\end{proof}

\Needspace{3\baselineskip}
\Needspace{8\baselineskip}
\phantomsection\label{app:weighted-rates}
\weightedrates*
\begin{proof}
Fix a player $i$. Apply \cref{lem:inverse-step-lipschitz} to bound $|\beta_i^{t+1}-\beta_i^t|$ by $|\hat{R}_i^t-\hat{R}_i^{t-1}|/\Lambda_d$. Bounding this soft regret increment by \cref{lem:soft-regret-decomposition} gives
\begin{equation}
 |\beta_i^{t+1}-\beta_i^t|
 \le\frac{\beta_i^t}{\Lambda_d}\bigl[\KL(x_i^t\|p_i^t)+\KL(x_i^t\|q_i^t)\bigr]
 +\frac{\log d_i}{\Lambda_d}|\beta_i^t-\beta_i^{t-1}|.
 \label{eq:rate-variation-recurrence}
\end{equation}
The term $(\log d_i/\Lambda_d)|\beta_i^t-\beta_i^{t-1}|$ on the right-hand side accounts for changes in the inverse step size and vanishes at $t=1$ since $\beta_i^0=\beta_i^1$. Multiply by $\Lambda_d/(\beta_i^t)^3$ and sum over $t=1,\ldots,T$. Reindexing the contribution $(\log d_i)\sum_{t=2}^T|\beta_i^t-\beta_i^{t-1}|/(\beta_i^t)^3$ and using the ratio bound in \cref{lem:rates} shows that it is at most $1/12$ of the left side below, since $(\log d_i/\Lambda_d)(21/20)^3<1/12$. Absorbing it and using $12/11\le3/2$ gives
\begin{equation}
 \Lambda_d\sum_{t=1}^{T}\frac{|\beta_i^{t+1}-\beta_i^t|}{(\beta_i^t)^3}
 \le\frac32\sum_{t=1}^{T}
 \frac{\KL(x_i^t\|p_i^t)+\KL(x_i^t\|q_i^t)}{(\beta_i^t)^2}.
 \label{eq:raw-weighted-rate-changes}
\end{equation}
For $T\ge2$, drop the zero $t=1$ term on the left of \eqref{eq:regularization-rate-changes} and reindex to obtain
\begin{align*}
 (\log d_i)\sum_{t=1}^{T}\frac{|\beta_i^t-\beta_i^{t-1}|}{(\beta_i^t)^3}
 &= (\log d_i)\sum_{t=1}^{T-1}\frac{|\beta_i^{t+1}-\beta_i^t|}{(\beta_i^{t+1})^3}\\*
 &\le (\log d_i)\left(\frac{21}{20}\right)^3
 \sum_{t=1}^{T-1}\frac{|\beta_i^{t+1}-\beta_i^t|}{(\beta_i^t)^3}\\*
 &\le \frac32\left(\frac{21}{20}\right)^3\frac{\log d_i}{\Lambda_d}
 \sum_{t=1}^{T-1}\frac{\KL(x_i^t\|p_i^t)+\KL(x_i^t\|q_i^t)}{(\beta_i^t)^2}\\*
 &\le \frac18\sum_{t=1}^{T}\frac{\KL(x_i^t\|p_i^t)+\KL(x_i^t\|q_i^t)}{(\beta_i^t)^2}.
\end{align*}
The first inequality uses $(\beta_i^t/\beta_i^{t+1})^3\le(21/20)^3$ from \cref{lem:rates}. The second applies \eqref{eq:raw-weighted-rate-changes} at horizon $T-1$. Since $(\log d_i)/\Lambda_d\le1/16$, the coefficient in the third line is below $1/8$. Using the nonnegativity of KL and adding the term at $t=T$, we obtain the upper bound in \eqref{eq:regularization-rate-changes}.

\Needspace{6\baselineskip}
To prove \eqref{eq:trajectory-rates}, we bound the squared increments of the inverse step sizes directly.
Write $\delta^t=|\beta_i^t-\beta_i^{t-1}|$ and $K^t=\KL(x_i^t\|p_i^t)+\KL(x_i^t\|q_i^t)$.
Let $c^t=\Phi_{\beta_i^t}(L_i^t)-\Phi_{\beta_i^t}(L_i^{t-1})$ be the change in the regularized minimum at time $t$, with $\Phi$ as in \cref{lem:entropy-minimum}.
The minimizers defining $\Phi_{\beta_i^t}(L_i^t)$ and $\Phi_{\beta_i^t}(L_i^{t-1})$ are $p_i^t$ and $q_i^t$, respectively. Using $p_i^t$ as a candidate in the second minimization and $q_i^t$ in the first gives
\begin{align*}
 \Phi_{\beta_i^t}(L_i^{t-1})
 &\le\ip{p_i^t}{L_i^{t-1}}+\beta_i^t\KL(p_i^t\|\one/d_i)
 =\Phi_{\beta_i^t}(L_i^t)-\ip{p_i^t}{\ell_i^t},\\*
 \Phi_{\beta_i^t}(L_i^t)
 &\le\ip{q_i^t}{L_i^t}+\beta_i^t\KL(q_i^t\|\one/d_i)
 =\Phi_{\beta_i^t}(L_i^{t-1})+\ip{q_i^t}{\ell_i^t}.
\end{align*}
The equalities use $L_i^t=L_i^{t-1}+\ell_i^t$. Rearranging gives $\ip{p_i^t}{\ell_i^t}\le c^t\le\ip{q_i^t}{\ell_i^t}$, and hence
\begin{align*}
 \ip{x_i^t}{\ell_i^t}-c^t
 &\le\ip{x_i^t-p_i^t}{\ell_i^t}
 \le\frac12\norm{x_i^t-p_i^t}_1
 \le\sqrt{\frac{\KL(x_i^t\|p_i^t)}2},\\*
 c^t-\ip{x_i^t}{\ell_i^t}
 &\le\ip{q_i^t-x_i^t}{\ell_i^t}
 \le\frac12\norm{x_i^t-q_i^t}_1
 \le\sqrt{\frac{\KL(x_i^t\|q_i^t)}2}.
\end{align*}
The middle inequalities use $\ell_i^t\in[0,1]^{d_i}$ and the fact that each difference of probability vectors sums to zero; the last inequalities are Pinsker's inequality.
If $\ip{x_i^t}{\ell_i^t}\ge c^t$, squaring the first line gives $(\ip{x_i^t}{\ell_i^t}-c^t)^2\le\KL(x_i^t\|p_i^t)/2\le K^t/2$. If $\ip{x_i^t}{\ell_i^t}<c^t$, squaring the second line gives the same bound with $q_i^t$ in place of $p_i^t$.
The decomposition in \cref{lem:soft-regret-decomposition} gives $\hat R_i^t-\hat R_i^{t-1}=\ip{x_i^t}{\ell_i^t}-c^t-a_i^t(\beta_i^t-\beta_i^{t-1})$.
Thus \cref{lem:inverse-step-lipschitz}, $a_i^t\le\log d_i$, and $(a+b)^2\le2a^2+2b^2$ imply
\begin{align*}
 \Lambda_d^2(\delta^{t+1})^2
 &\le|\hat R_i^t-\hat R_i^{t-1}|^2\\*
 &\le2(\ip{x_i^t}{\ell_i^t}-c^t)^2+2(\log d_i)^2(\delta^t)^2
 \le K^t+2(\log d_i)^2(\delta^t)^2.
\end{align*}
For $T\ge2$, divide by $(\beta_i^t)^2$ and sum over $t=1,\ldots,T-1$.
Since $\delta^1=0$, reindexing the preceding increments and using the ratio bound in \cref{lem:rates} gives
\begin{align*}
 \Lambda_d^2\sum_{t=1}^{T-1}\frac{(\delta^{t+1})^2}{(\beta_i^t)^2}
 &\le\sum_{t=1}^{T-1}\frac{K^t}{(\beta_i^t)^2}
   +2(\log d_i)^2\sum_{t=1}^{T-2}\frac{(\delta^{t+1})^2}{(\beta_i^{t+1})^2}\\*
 &\le\sum_{t=1}^{T-1}\frac{K^t}{(\beta_i^t)^2}
   +2\left(\frac{21}{320}\right)^2\Lambda_d^2
      \sum_{t=1}^{T-1}\frac{(\delta^{t+1})^2}{(\beta_i^t)^2}.
\end{align*}
The last line uses $(\log d_i)/\Lambda_d\le1/16$ and extends a nonnegative sum by one term.
Absorb the last sum into the left side.
Finally, $|\log(a/b)|\le|a-b|/\min\{a,b\}$ and the same ratio bound give
\begin{align*}
 \Lambda_d^2\sum_{t=1}^{T-1}\left(\log\frac{\beta_i^{t+1}}{\beta_i^t}\right)^2
 &\le\left(\frac{21}{20}\right)^2\Lambda_d^2
      \sum_{t=1}^{T-1}\frac{(\delta^{t+1})^2}{(\beta_i^t)^2}\\*
 &\le\frac{(21/20)^2}{1-2(21/320)^2}
      \sum_{t=1}^{T-1}\frac{K^t}{(\beta_i^t)^2}
 \le2\sum_{t=1}^{T}\frac{K^t}{(\beta_i^t)^2}.
\end{align*}
The last inequality uses $(21/20)^2/[1-2(21/320)^2]<2$ and $K^T\ge0$.
For $T=1$, the term in \eqref{eq:regularization-rate-changes} vanishes because $\beta_i^0=\beta_i^1$, and the sum in \eqref{eq:trajectory-rates} is empty.
\end{proof}

\Needspace{6\baselineskip}
\section{Proof of Self-Play Prediction Error Bound}\label{app:prediction-proof}

In this section, we prove the two main lemmas, \Cref{lem:reduction} and \Cref{lem:loss-differences}, from \Cref{sec:prediction},
using the following helper lemmas.

\Needspace{8\baselineskip}
\phantomsection\label{app:kl-bounds}

\begin{lemma}[Variance bounds on the KL divergences]\label{lem:kl-bounds}
For Optimistic Hedge with any positive inverse step sizes $\beta_i^t$ and an arbitrary loss sequence $\{\ell_i^t\}_{t\ge1}\subseteq[0,1]^{d_i}$, every $t\ge1$ satisfies $\KL(x_i^t\|p_i^t)\le1/[2(\beta_i^t)^2]$ and $\KL(x_i^t\|q_i^t)\le1/[8(\beta_i^t)^2]$.
Moreover, if $\beta_i^t\ge4$, then
\begin{align}
 \frac{\Var_{x_i^t}(\Delta\ell_i^t)}{4(\beta_i^t)^2}
 &\le\KL(x_i^t\|p_i^t)\le\frac{\Var_{x_i^t}(\Delta\ell_i^t)}{(\beta_i^t)^2},\label{eq:prediction-kl-variance}\\
 \frac{\Var_{x_i^t}(\ell_i^{t-1})}{4(\beta_i^t)^2}
 &\le\KL(x_i^t\|q_i^t)\le\frac{\Var_{x_i^t}(\ell_i^{t-1})}{(\beta_i^t)^2}.
 \label{eq:kl-variance}
\end{align}
At the first round, $\KL(x_i^1\|p_i^1)\le1/(8\beta^2)$ and $\KL(x_i^1\|q_i^1)=0$.
\end{lemma}
\begin{proof}
Optimistic Hedge satisfies $x_i^t=\softmax(-(L_i^{t-1}+\ell_i^{t-1})/\beta_i^t)$. Therefore, we have
\begin{align*}
 p_i^t&=\softmax\!\left(\log x_i^t-\frac{\Delta\ell_i^t}{\beta_i^t}\right),\\*
 q_i^t&=\softmax\!\left(\log x_i^t+\frac{\ell_i^{t-1}}{\beta_i^t}\right).
\end{align*}
The log normalizer in $\log x_i^t$ is common to all coordinates and therefore cancels under softmax.
Because $\Delta\ell_i^t\in[-1,1]^{d_i}$ and $\ell_i^{t-1}\in[0,1]^{d_i}$, the first bound in \cref{lem:reweighting-kl} gives
\begin{align*}
 \KL(x_i^t\|p_i^t)&\le\frac{\operatorname{osc}(\Delta\ell_i^t)^2}{8(\beta_i^t)^2}\le\frac{1}{2(\beta_i^t)^2},\\*
 \KL(x_i^t\|q_i^t)&\le\frac{\operatorname{osc}(\ell_i^{t-1})^2}{8(\beta_i^t)^2}\le\frac{1}{8(\beta_i^t)^2}.
\end{align*}
These bounds hold for every positive $\beta_i^t$.

If $\beta_i^t\ge4$, the vectors $-\Delta\ell_i^t/\beta_i^t$ and $\ell_i^{t-1}/\beta_i^t$ have oscillations at most $2/\beta_i^t\le1/2$ and $1/\beta_i^t\le1/4$, respectively.
Applying the variance bounds in \cref{lem:reweighting-kl} with $u=x_i^t$ and these two choices of $z$, and using $\Var_u(cz)=c^2\Var_u(z)$, gives \eqref{eq:prediction-kl-variance} and \eqref{eq:kl-variance}.
Finally, at $t=1$, we have $\ell_i^0=0$ and $\beta_i^1=\beta$, so $\Delta\ell_i^1=\ell_i^1$ has oscillation at most $1$.
The first bound above then improves to $1/(8\beta^2)$, while $q_i^1=x_i^1=\one/d_i$ gives the zero divergence.
\end{proof}

\begin{lemma}[Strategy movement bound]\label{lem:movement}
For player $i$, fix arbitrary strategies $\{x_i^t\}_{t\ge1}$ and an arbitrary loss sequence $\{\ell_i^t\}_{t\ge1}\subseteq[0,1]^{d_i}$. Suppose $\beta_i^1=\beta\ge4$ and the step sizes follow $\beta_i^{t+1}=\sqrt{\beta^2+(\hat{R}_i^t/\Lambda_d)^2}$ for every $t\ge1$. Then, for every $T\ge1$,
\begin{equation}
 8\sum_{t=2}^T\left\|
 \frac{\sqrt{x_i^t}}{\beta_i^t}-
 \frac{\sqrt{x_i^{t-1}}}{\beta_i^{t-1}}\right\|_2^2
 \le\frac{13}{\beta^2}\sum_{t=1}^T\KL(x_i^t\|p_i^t)
    +51\sum_{t=1}^T\frac{\KL(x_i^t\|q_i^t)}{(\beta_i^t)^2}.
 \label{eq:trajectory-movement}
\end{equation}
\end{lemma}
\begin{proof}
We compare $\sqrt{x_i^{t-1}}/\beta_i^{t-1}$ and $\sqrt{x_i^t}/\beta_i^t$ through $\sqrt{p_i^{t-1}}/\beta_i^{t-1}$ and $\sqrt{q_i^t}/\beta_i^t$. Since $p_i^{t-1}$ and $q_i^t$ both use $L_i^{t-1}$, we have $q_i^t=\softmax((\beta_i^{t-1}/\beta_i^t)\log p_i^{t-1})$. Hence, we have
\begin{align}
 \left\|\frac{\sqrt{p_i^{t-1}}}{\beta_i^{t-1}}-\frac{\sqrt{q_i^t}}{\beta_i^t}\right\|_2
 &\le \left|\frac1{\beta_i^{t-1}}-\frac1{\beta_i^t}\right|
      +\frac{\|\sqrt{p_i^{t-1}}-\sqrt{q_i^t}\|_2}{\beta_i^t}\notag\\*
 &\le \frac{1+(1+\log d_i)/\sqrt2}{\beta}
      \left|\log\frac{\beta_i^t}{\beta_i^{t-1}}\right|
 \le \frac{\Lambda_d}{8\beta}\left|\log\frac{\beta_i^t}{\beta_i^{t-1}}\right|.
 \label{eq:movement}
\end{align}
The first inequality uses $\|\sqrt{p_i^{t-1}}\|_2=1$. For the second, we apply the $1/\beta$-Lipschitz bound for $z\mapsto e^{-z}$ on $[\log\beta,\infty)$ at $\log\beta_i^{t-1}$ and $\log\beta_i^t$, and apply \cref{lem:rescaling} to $\|\sqrt{p_i^{t-1}}-\sqrt{q_i^t}\|_2$. The last inequality uses $\Lambda_d=16(1+\log d)$ and $d_i\le d$.

By the triangle inequality,
\[
 \left\|\frac{\sqrt{x_i^{t-1}}}{\beta_i^{t-1}}-\frac{\sqrt{x_i^t}}{\beta_i^t}\right\|_2
 \le \frac{\|\sqrt{x_i^{t-1}}-\sqrt{p_i^{t-1}}\|_2}{\beta_i^{t-1}}
      +\left\|\frac{\sqrt{p_i^{t-1}}}{\beta_i^{t-1}}-\frac{\sqrt{q_i^t}}{\beta_i^t}\right\|_2
      +\frac{\|\sqrt{q_i^t}-\sqrt{x_i^t}\|_2}{\beta_i^t}.
\]
We apply $(a+b+c)^2\le\tfrac32a^2+6b^2+6c^2$ (Cauchy--Schwarz) to the right-hand side, using \eqref{eq:kl-distance} for the first and third summands and \eqref{eq:movement} for the second. Multiplying by $8$ and summing over $2\le t\le T$ gives
\begin{align*}
 8\sum_{t=2}^T\left\|\frac{\sqrt{x_i^t}}{\beta_i^t}
                  -\frac{\sqrt{x_i^{t-1}}}{\beta_i^{t-1}}\right\|_2^2
 &\le 12\sum_{t=1}^{T-1}\frac{\KL(x_i^t\|p_i^t)}{(\beta_i^t)^2}
      +48\sum_{t=2}^T\frac{\KL(x_i^t\|q_i^t)}{(\beta_i^t)^2}\\*
 &\quad+\frac{3\Lambda_d^2}{4\beta^2}\sum_{t=2}^T
       \left(\log\frac{\beta_i^t}{\beta_i^{t-1}}\right)^2.
\end{align*}
By \eqref{eq:trajectory-rates} and $3/(2\beta^2)\le1$, the last term is at most $\sum_{t=1}^T[\KL(x_i^t\|p_i^t)+\KL(x_i^t\|q_i^t)]/(\beta_i^t)^2$. Extending the first KL sum to $t=T$ and the second to $t=1$ therefore gives coefficients $12+1=13$ and $48+1=49$, respectively. Using $\beta_i^t\ge\beta$ in the first sum and $49\le51$ in the second proves \eqref{eq:trajectory-movement}.
\end{proof}

\Needspace{10\baselineskip}
\begin{lemma}[Weighted sum of the first difference variance]\label{lem:weighted-finite-differences}
Fix integers $k\ge1$ and $T\ge2$, probability vectors $x^1,\ldots,x^T\in\Delta_k$, vectors $\ell^1,\ldots,\ell^T\in[0,1]^k$, and nonnegative weights $w^2,\ldots,w^T$. Write $\Delta\ell^t=\ell^t-\ell^{t-1}$ for $2\le t\le T$ and $\Delta^2\ell^t=\Delta\ell^{t+1}-\Delta\ell^t$ for $2\le t\le T-1$. Let $\Cov_x(u,v)=\sum_{h\in[k]}x_h(u_h-\ip xu)(v_h-\ip xv)$ denote the covariance under $x$ for two vectors $u,v\in\mathbb{R}^k$. Then
\begin{align}
 \sum_{t=2}^T(w^t)^2\Var_{x^t}(\Delta\ell^t)
 &\le(w^2)^2+(w^T)^2
 +2\left|\sum_{t=2}^{T-1}(w^{t+1})^2\Cov_{x^{t+1}}(\ell^t,\Delta^2\ell^t)\right|\nonumber\\*
 &\quad+8\sum_{t=2}^{T-1}
 \|w^{t+1}\sqrt{x^{t+1}}-w^t\sqrt{x^t}\|_2^2.
 \label{eq:weighted-finite-differences}
\end{align}
\end{lemma}
\begin{proof}
Write $C^t(u,v)=(w^t)^2\Cov_{x^t}(u,v)$. Summation by parts gives
\begin{align}
 &\sum_{t=2}^T(w^t)^2\Var_{x^t}(\Delta\ell^t)=\sum_{t=2}^T C^t(\Delta\ell^t,\Delta\ell^t)\nonumber\\
 &=\sum_{t=2}^T C^t(\ell^t,\Delta\ell^t)
   -\sum_{t=2}^T C^t(\ell^{t-1},\Delta\ell^t)\nonumber\\
 &=C^T(\ell^T,\Delta\ell^T)-C^2(\ell^1,\Delta\ell^2)
   +\sum_{t=2}^{T-1}\bigl[C^t(\ell^t,\Delta\ell^t)-C^{t+1}(\ell^t,\Delta\ell^{t+1})\bigr]\nonumber\\
 &=C^T(\ell^T,\Delta\ell^T)-C^2(\ell^1,\Delta\ell^2)
   -\sum_{t=2}^{T-1}C^{t+1}(\ell^t,\Delta^2\ell^t)
   +\sum_{t=2}^{T-1}(C^t-C^{t+1})(\ell^t,\Delta\ell^t).
 \label{eq:covariance-summation}
\end{align}
The second equality reindexes the second sum and separates the terms at $t=2$ and $t=T$. The last adds and subtracts $C^{t+1}(\ell^t,\Delta\ell^t)$ inside the sum and uses $\Delta^2\ell^t=\Delta\ell^{t+1}-\Delta\ell^t$. The first two terms on the last line of \eqref{eq:covariance-summation} satisfy
\begin{align*}
 |C^T(\ell^T,\Delta\ell^T)|+|C^2(\ell^1,\Delta\ell^2)|
 &\le (w^T)^2\sqrt{\Var_{x^T}(\ell^T)\Var_{x^T}(\Delta\ell^T)}\\*
 &\quad+(w^2)^2\sqrt{\Var_{x^2}(\ell^1)\Var_{x^2}(\Delta\ell^2)}
 \le\frac{(w^2)^2+(w^T)^2}{2}.
\end{align*}
The first inequality is Cauchy--Schwarz; the second uses that losses and their differences have standard deviations at most $1/2$ and $1$, respectively.

To control the last term in \eqref{eq:covariance-summation}, fix $2\le t\le T-1$ and set $x=x^t$, $y=x^{t+1}$, and $\delta=w^t\sqrt x-w^{t+1}\sqrt y$. For $u\in[0,1]^k$ and $v\in[-1,1]^k$,
\begin{align}
 |(C^t-C^{t+1})(u,v)|
 &=\big|(w^t)^2\Cov_x(u,v)-(w^{t+1})^2\Cov_y(u,v)\big|\nonumber\\*
 &=\left|\sum_h\big[(w^t)^2x_h-(w^{t+1})^2y_h\big]
 (u_h-\ip yu)(v_h-\ip xv)\right|\nonumber\\*
 &=\left|\sum_h(2w^t\sqrt{x_h}\delta_h-\delta_h^2)
 (u_h-\ip yu)(v_h-\ip xv)\right|\nonumber\\*
 &\le 2\|\delta\|_2^2+2w^t\left|\sum_h(\sqrt{x_h}\delta_h)
 (u_h-\ip yu)(v_h-\ip xv)\right| \nonumber\\*
 &\le2\|\delta\|_2^2+2w^t\|\delta\|_2\sqrt{\Var_x(v)}
 \tag{Cauchy--Schwarz}\\*
 &\le\frac12C^t(v,v)+4\|\delta\|_2^2.
 \label{eq:covariance-weight-change}
\end{align}
The second equality uses $\sum_h x_h(v_h-\ip xv)=\sum_h y_h(u_h-\ip yu)=0$ and replaces $\ip xu$ by $\ip yu$ in $\Cov_x(u,v)$, and $\ip yv$ by $\ip xv$ in $\Cov_y(u,v)$. The third equality uses $w^{t+1}\sqrt{y_h}=w^t\sqrt{x_h}-\delta_h$. The first inequality uses $|u_h-\ip yu|\le1$ and $|v_h-\ip xv|\le2$. The last uses $2rs\le2r^2+s^2/2$ with $r=\|\delta\|_2$ and $s=w^t\sqrt{\Var_x(v)}$.

Write $V=\sum_{t=2}^T C^t(\Delta\ell^t,\Delta\ell^t)$. Applying \eqref{eq:covariance-weight-change} with $u=\ell^t$ and $v=\Delta\ell^t$ bounds the last sum of \eqref{eq:covariance-summation} as
\[
 \sum_{t=2}^{T-1}(C^t-C^{t+1})(\ell^t,\Delta\ell^t)
 \le\underbrace{\frac12\sum_{t=2}^{T-1}C^t(\Delta\ell^t,\Delta\ell^t)}_{[V-C^T(\Delta\ell^T,\Delta\ell^T)]/2}
 +4\sum_{t=2}^{T-1}\|w^{t+1}\sqrt{x^{t+1}}-w^t\sqrt{x^t}\|_2^2.
\]
Substituting this bound into \eqref{eq:covariance-summation}, moving $V/2$ to the left, using the estimate $|C^T(\ell^T,\Delta\ell^T)|+|C^2(\ell^1,\Delta\ell^2)|\le \tfrac12[(w^2)^2+(w^T)^2]$, and multiplying by two gives
\[
 V+C^T(\Delta\ell^T,\Delta\ell^T)
 \le (w^2)^2+(w^T)^2
 +2\left|\sum_{t=2}^{T-1}C^{t+1}(\ell^t,\Delta^2\ell^t)\right|
 +8\sum_{t=2}^{T-1}\|w^{t+1}\sqrt{x^{t+1}}-w^t\sqrt{x^t}\|_2^2.
\]
Since $C^T(\Delta\ell^T,\Delta\ell^T)\ge0$, this proves \eqref{eq:weighted-finite-differences}.
\end{proof}

\Needspace{5\baselineskip}
\reductionlemma*
\begin{proof}
For $T=1$, \cref{lem:kl-bounds} gives $P_i(1)\le1/(8\beta^2)$, so assume $T\ge2$.
We first prove
\begin{equation}
 \left(1-\frac{13}{\beta^2}\right)P_i(T)
 \le\frac{17}{8\beta^2}+51Q_i(T)
 +2\left|\sum_{t=2}^{T-1}
 \frac{\Cov_{x_i^{t+1}}(\ell_i^t,\Delta^2\ell_i^t)}{(\beta_i^{t+1})^2}\right|.
 \label{eq:covariance-reduction}
\end{equation}
Taking $x^t=x_i^t$, $\ell^t=\ell_i^t$, and $w^t=1/\beta_i^t$ in \eqref{eq:weighted-finite-differences}, we obtain
\begin{align*}
 P_i(T)
 &\le \frac1{8\beta^2}
      +\sum_{t=2}^T\frac{\Var_{x_i^t}(\Delta\ell_i^t)}{(\beta_i^t)^2}
      \tag{\Cref{lem:kl-bounds}}\\*
 &\le \underbrace{\frac1{8\beta^2}+\frac2{\beta^2}}_{17/(8\beta^2)}
    +8\sum_{t=2}^{T-1}\left\|
        \frac{\sqrt{x_i^{t+1}}}{\beta_i^{t+1}}-
        \frac{\sqrt{x_i^t}}{\beta_i^t}\right\|_2^2
    +2\left|\sum_{t=2}^{T-1}
      \frac{\Cov_{x_i^{t+1}}(\ell_i^t,\Delta^2\ell_i^t)}{(\beta_i^{t+1})^2}\right|
      \tag{\Cref{lem:weighted-finite-differences}}\\*
 &\le \frac{17}{8\beta^2}+\frac{13}{\beta^2}P_i(T)+51Q_i(T)
    +2\left|\sum_{t=2}^{T-1}
      \frac{\Cov_{x_i^{t+1}}(\ell_i^t,\Delta^2\ell_i^t)}{(\beta_i^{t+1})^2}\right|.
      \tag{\Cref{lem:movement}}
\end{align*}
The first term in the second inequality uses $(w^2)^2+(w^T)^2\le2/\beta^2$, which follows from $\beta_i^2,\beta_i^T\ge\beta$. Collecting the two terms in $P_i(T)$ gives \eqref{eq:covariance-reduction}.

We now bound the last term in \eqref{eq:covariance-reduction}. Applying Cauchy--Schwarz first under $x_i^{t+1}$ and then over rounds gives
\begin{align*}
 2\left|\sum_{t=2}^{T-1}
      \frac{\Cov_{x_i^{t+1}}(\ell_i^t,\Delta^2\ell_i^t)}{(\beta_i^{t+1})^2}\right|
 &\le2\sum_{t=2}^{T-1}
      \frac{\sqrt{\Var_{x_i^{t+1}}(\ell_i^t)\Var_{x_i^{t+1}}(\Delta^2\ell_i^t)}}{(\beta_i^{t+1})^2}\\*
 &\le2\left(\sum_{t=2}^{T-1}
      \frac{\Var_{x_i^{t+1}}(\ell_i^t)}{(\beta_i^{t+1})^4}\right)^{1/2}
      D_i(T).
\end{align*}
The second inequality uses $\Var_{x_i^{t+1}}(\Delta^2\ell_i^t)\le\|\Delta^2\ell_i^t\|_\infty^2$.
To bound the remaining variance sum, observe that $\ell_i^t$ is the loss used to predict round $t+1$. Thus \eqref{eq:kl-variance} at round $t+1$ gives
\[
 \sum_{t=2}^{T-1}\frac{\Var_{x_i^{t+1}}(\ell_i^t)}{(\beta_i^{t+1})^4}
 \le 4\sum_{s=3}^{T}\frac{\KL(x_i^s\|q_i^s)}{(\beta_i^s)^2}
 \le 4Q_i(T).
\]
The covariance term in \eqref{eq:covariance-reduction} is therefore at most $4\sqrt{Q_i(T)}\,D_i(T)$, which proves \eqref{eq:reduction}.
\end{proof}

\Needspace{5\baselineskip}
Let $r_i^t$ be the strategy player $i$ would use next if its inverse step size remained at $\beta_i^t$. Since $L_i^t=L_i^{t-1}+\ell_i^t$, $p_i^t=\softmax(-L_i^t/\beta_i^t)$ can be written as the normalized geometric mean of $q_i^t=\softmax(-L_i^{t-1}/\beta_i^t)$ and $r_i^t$: 
\begin{equation}
 r_i^t=\softmax(-(L_i^t+\ell_i^t)/\beta_i^t),
 \qquad
 p_{i,a}^t=\frac{\sqrt{q_{i,a}^t r_{i,a}^t}}
 {\sum_b\sqrt{q_{i,b}^t r_{i,b}^t}}.
 \label{eq:geometric-mean}
\end{equation}

\noindent Also recall the definition of $\ell_i(x_{-i})$ from \eqref{eq:game}:
\[
[\ell_i(x_{-i})]_{a_i}
 \coloneqq\sum_{a_{-i}\in \prod_{j\in[n]\setminus\{i\}} [d_j]}
 \ell_i(a_i,a_{-i})\prod_{j\in[n]\setminus\{i\}}x_{j,a_j}.
\]

\begin{lemma}[Second-order loss difference bound]\label{lem:fixed-rate}
Fix player $i$ and round $t\ge1$. If $\beta_j^t\ge4$ for every $j\ne i$, then
\begin{equation}
 \norm{\ell_i(r_{-i}^t)-2\ell_i(p_{-i}^t)+\ell_i(q_{-i}^t)}_\infty
 \le10\sum_{j\ne i}\bigl[\KL(x_j^t\|p_j^t)+\KL(x_j^t\|q_j^t)\bigr].
 \label{eq:geometric-mean-loss-bound}
\end{equation}
\end{lemma}
\begin{proof}
For positive $q,r\in\Delta_k$, let $p_a=\sqrt{q_ar_a}/\sum_b\sqrt{q_br_b}$. For every $h\in[0,1]^k$, using the identity $\sum_a \sqrt{q_a r_a}(h_a - \ip{p}{h})=0$ one can verify
\begin{equation}
 \ip{q+r-2p}{h}
 =\sum_a(\sqrt{q_a}-\sqrt{r_a})^2(h_a-\ip ph),
 \label{eq:geometric-mean-identity}
\end{equation}
and thus $|\ip{q+r-2p}{h}|\le\norm{\sqrt q-\sqrt r}_2^2$.
By \eqref{eq:geometric-mean}, the product distributions $Q=\prod_{j\ne i}q_j^t$ and $R=\prod_{j\ne i}r_j^t$ have normalized geometric mean $\prod_{j\ne i}p_j^t$. Applying the preceding inequality to $Q$ and $R$ and using \cref{lem:product-comparisons} gives
\[
 \norm{\ell_i(r_{-i}^t)-2\ell_i(p_{-i}^t)+\ell_i(q_{-i}^t)}_\infty
 \le\norm{\sqrt Q-\sqrt R}_2^2
 \le\sum_{j\ne i}\norm{\sqrt{q_j^t}-\sqrt{r_j^t}}_2^2.
\]

To bound the sum on the right-hand side, fix an opponent $j$ and write $q=q_j^t$, $r=r_j^t$, $p=p_j^t$, $x=x_j^t$, and $\ell=\ell_j^t$.
Bounded losses give $p_a/q_a=e^{-\ell_a/\beta_j^t}/\sum_bq_b e^{-\ell_b/\beta_j^t}\le e^{1/\beta_j^t}$. By positivity and Cauchy--Schwarz, $0<\sum_a\sqrt{q_ar_a}\le1$. Using $2(1-z)\le z^{-2}-1$ for $0<z\le1$ and $p_a=\sqrt{q_ar_a}/\sum_b\sqrt{q_br_b}$ from \eqref{eq:geometric-mean}, we have
\begin{align*}
 \norm{\sqrt q-\sqrt r}_2^2
 &=2\left(1-\sum_a\sqrt{q_ar_a}\right)
 \le\left(\sum_a\sqrt{q_ar_a}\right)^{-2}-1
 =\sum_a\frac{(p_a-q_a)^2}{q_a}\\
 &\le(1+e^{1/(2\beta_j^t)})^2\norm{\sqrt p-\sqrt q}_2^2
 \le10\bigl[\KL(x\|p)+\KL(x\|q)\bigr].
\end{align*}
Here $\beta_j^t\ge4$, and the triangle inequality through $\sqrt x$ and \eqref{eq:kl-distance} give $\|\sqrt p-\sqrt q\|_2^2\le2[\KL(x\|p)+\KL(x\|q)]$. Also, $2(1+e^{1/8})^2<10$.
Summing over opponents proves the lemma.
\end{proof}

\phantomsection\label{app:loss-differences}
\losslemma*
\begin{proof}
The result is immediate for $T\le2$. Fix $T\ge3$, a player $i$, and an interior round $2\le t\le T-1$. To compare the strategies with fixed inverse step sizes in \cref{lem:fixed-rate} with the played strategies, decompose
\begin{align*}
 \Delta^2\ell_i^t
 ={}&\bigl[\ell_i(r_{-i}^t)-2\ell_i(p_{-i}^t)+\ell_i(q_{-i}^t)\bigr]\\
 &+\bigl[\ell_i^{t+1}-\ell_i(r_{-i}^t)\bigr]
 -2\bigl[\ell_i^t-\ell_i(p_{-i}^t)\bigr]
 +\bigl[\ell_i^{t-1}-\ell_i(q_{-i}^t)\bigr].
\end{align*}
The sup norm of the first bracket is also at most $2$ since its loss vectors have entries in $[0,1]$.
By assumption, $\ell_i^s=\ell_i(x_{-i}^s)$. Using $h=\ell_i(a_i,\cdot)$ for each action $a_i$ in \cref{lem:product-comparisons} gives
\begin{align*}
 \|\Delta^2\ell_i^t\|_\infty
 &\le \min\left\{2,\,10\sum_{j\ne i}[\KL(x_j^t\|p_j^t)+\KL(x_j^t\|q_j^t)]\right\}
 \tag{Triangle inequality, \Cref{lem:fixed-rate}}\\*
 &\quad+\left(\sum_{j\ne i}\norm{\sqrt{x_j^{t+1}}-\sqrt{r_j^t}}_2^2\right)^{1/2}
 +2\left(\sum_{j\ne i}\norm{\sqrt{x_j^t}-\sqrt{p_j^t}}_2^2\right)^{1/2}\\*
 &\quad+\left(\sum_{j\ne i}\norm{\sqrt{x_j^{t-1}}-\sqrt{q_j^t}}_2^2\right)^{1/2}.
 \tag{\Cref{lem:product-comparisons}}
\end{align*}

For each opponent $j$, $x_j^{t+1}$ and $r_j^t$ use the same loss vector $L_j^t+\ell_j^t$, and $q_j^t$ and $p_j^{t-1}$ use $L_j^{t-1}$. By their definitions,
\begin{align*}
 x_j^{t+1}
 &=\softmax\!\left(-\frac{L_j^t+\ell_j^t}{\beta_j^{t+1}}\right)
 =\softmax\!\left(\frac{\beta_j^t}{\beta_j^{t+1}}\log r_j^t\right),\\*
 q_j^t
 &=\softmax\!\left(-\frac{L_j^{t-1}}{\beta_j^t}\right)
 =\softmax\!\left(\frac{\beta_j^{t-1}}{\beta_j^t}\log p_j^{t-1}\right).
\end{align*}
The second equality in each line uses $\log\softmax(z)=z-\log(\sum_a e^{z_a})\one$ and $\softmax(z+c\one)=\softmax(z)$ for any scalar $c$.
Taking square roots on \eqref{eq:rescaling-distance} in \cref{lem:rescaling} and using $1+\log d_j\le\Lambda_d/16$ bounds the effect of each change in inverse step size, yielding
\begin{align*}
 \norm{\sqrt{x_j^{t+1}}-\sqrt{r_j^t}}_2
 &\le\frac{\Lambda_d}{16\sqrt2}\left|\log\frac{\beta_j^{t+1}}{\beta_j^t}\right|,
 \tag{\Cref{lem:rescaling}}\\*
 \norm{\sqrt{x_j^{t-1}}-\sqrt{q_j^t}}_2
 &\le\norm{\sqrt{x_j^{t-1}}-\sqrt{p_j^{t-1}}}_2
 +\norm{\sqrt{p_j^{t-1}}-\sqrt{q_j^t}}_2
 \tag{Triangle inequality}\\*
 &\le\sqrt{\KL(x_j^{t-1}\|p_j^{t-1})}
 +\frac{\Lambda_d}{16\sqrt2}\left|\log\frac{\beta_j^t}{\beta_j^{t-1}}\right|.
 \tag{\Cref{lem:probability-comparisons}, \Cref{lem:rescaling}}
\end{align*}
Substituting the last bound into the sum over opponents gives
\begin{align*}
 \left(\sum_{j\ne i}\norm{\sqrt{x_j^{t-1}}-\sqrt{q_j^t}}_2^2\right)^{1/2}
 &\le\left(\sum_{j\ne i}
 \left[\sqrt{\KL(x_j^{t-1}\|p_j^{t-1})}
 +\frac{\Lambda_d}{16\sqrt2}\left|\log\frac{\beta_j^t}{\beta_j^{t-1}}\right|\right]^2\right)^{1/2}\\*
 &\le\left(\sum_{j\ne i}\KL(x_j^{t-1}\|p_j^{t-1})\right)^{1/2}
 +\frac{\Lambda_d}{16\sqrt2}
 \left(\sum_{j\ne i}\left(\log\frac{\beta_j^t}{\beta_j^{t-1}}\right)^2\right)^{1/2}.
 \tag{Euclidean triangle inequality}
\end{align*}
Using these estimates and $\|\sqrt{x_j^t}-\sqrt{p_j^t}\|_2\le\sqrt{\KL(x_j^t\|p_j^t)}$ from \cref{lem:probability-comparisons} in the loss bound gives
\begin{align}
 \|\Delta^2\ell_i^t\|_\infty
 &\le \underbrace{2\sqrt{\sum_{j\ne i}\KL(x_j^t\|p_j^t)}
       +\sqrt{\sum_{j\ne i}\KL(x_j^{t-1}\|p_j^{t-1})}}_{A_t}\nonumber\\*
 &\quad+\underbrace{\min\left\{2,\,10\sum_{j\ne i}
       [\KL(x_j^t\|p_j^t)+\KL(x_j^t\|q_j^t)]\right\}}_{B_t}\nonumber\\*
 &\quad+\underbrace{\frac{\Lambda_d}{16}
       \left(\sum_{j\ne i}\left[
       \left(\log\frac{\beta_j^t}{\beta_j^{t-1}}\right)^2+
       \left(\log\frac{\beta_j^{t+1}}{\beta_j^t}\right)^2\right]\right)^{1/2}}_{C_t}.
 \label{eq:pointwise-loss-difference}
\end{align}
The first term $A_t$ uses $\|\sqrt{x_j^s}-\sqrt{p_j^s}\|_2^2\le\KL(x_j^s\|p_j^s)$ at $s=t$ and $s=t-1$, and the second term $B_t$ uses the smaller of the KL bound in \eqref{eq:geometric-mean-loss-bound} and the bound $2$ from losses in $[0,1]$. The last term $C_t$ uses $\sqrt u+\sqrt v\le\sqrt{2(u+v)}$ with $u=\sum_{j\ne i}(\log(\beta_j^t/\beta_j^{t-1}))^2$ and $v=\sum_{j\ne i}(\log(\beta_j^{t+1}/\beta_j^t))^2$.

The triangle inequality over $t=2,\ldots,T-1$ gives
\begin{align*}
 \left(\sum_{t=2}^{T-1}A_t^2\right)^{1/2}
 &\le 2\left(\sum_{j\ne i}\sum_{t=2}^{T-1}\KL(x_j^t\|p_j^t)\right)^{1/2}
 +\left(\sum_{j\ne i}\sum_{s=1}^{T-2}\KL(x_j^s\|p_j^s)\right)^{1/2}\\*
 &\le 3\sqrt{\sum_{j\ne i}P_j(T)}.
\end{align*}
For $B_t$, apply $\min\{2,u+v\}\le\min\{2,u\}+v\le\sqrt{2u}+v$ with $u=10\sum_{j\ne i}\KL(x_j^t\|p_j^t)$ and $v=10\sum_{j\ne i}\KL(x_j^t\|q_j^t)$ to obtain
\begin{equation}
 B_t\le \sqrt{20\sum_{j\ne i}\KL(x_j^t\|p_j^t)}
            +10\sum_{j\ne i}\KL(x_j^t\|q_j^t).
 \label{eq:squared-kl}
\end{equation}
Using the triangle inequality over $t=2,\ldots,T-1$ in \eqref{eq:squared-kl} gives
\begin{align*}
 \left(\sum_{t=2}^{T-1} B_t^2\right)^{1/2}
 &\le \sqrt{20}\sqrt{\sum_{j\ne i}P_j(T)}
     +10\sum_{j\ne i}\left(\sum_{t=2}^{T-1}\KL(x_j^t\|q_j^t)^2\right)^{1/2}\\
 &\le \sqrt{20}\sqrt{\sum_{j\ne i}P_j(T)}
     +\frac{10}{\sqrt8}\sum_{j\ne i}\sqrt{Q_j(T)} .
\end{align*}
The second inequality uses $\KL(x_j^t\|q_j^t)^2\le\KL(x_j^t\|q_j^t)/[8(\beta_j^t)^2]$ by \cref{lem:kl-bounds}.

Summing $C_t^2$ over $2\le t\le T-1$ and setting $s=t-1$ in the first sum and $s=t$ in the second gives
\begin{align*}
 \sum_{t=2}^{T-1} C_t^2
 &=\frac{\Lambda_d^2}{256}\sum_{j\ne i}\left[
     \sum_{s=1}^{T-2}\left(\log\frac{\beta_j^{s+1}}{\beta_j^s}\right)^2
    +\sum_{s=2}^{T-1}\left(\log\frac{\beta_j^{s+1}}{\beta_j^s}\right)^2\right]\\*
 &\le \frac{\Lambda_d^2}{128}
     \sum_{j\ne i}\sum_{s=1}^{T-1}\left(\log\frac{\beta_j^{s+1}}{\beta_j^s}\right)^2
 \le\frac1{64}\sum_{j\ne i}\left(\frac{P_j(T)}{\beta^2}+Q_j(T)\right).
\end{align*}
The first inequality extends both sums to $1\le s\le T-1$. The second uses \eqref{eq:trajectory-rates} and $\beta_j^s\ge\beta$.
Taking square roots gives $(\sum_{t=2}^{T-1}C_t^2)^{1/2}\le(8\beta)^{-1}\sqrt{\sum_{j\ne i}P_j(T)}+\tfrac18\sum_{j\ne i}\sqrt{Q_j(T)}$. The triangle inequality in \eqref{eq:pointwise-loss-difference} now yields
\begin{align*}
 D_i(T)
 &\le\left(\sum_{t=2}^{T-1}A_t^2\right)^{1/2}
     +\left(\sum_{t=2}^{T-1}B_t^2\right)^{1/2}
     +\left(\sum_{t=2}^{T-1}C_t^2\right)^{1/2}\\*
 &\le \left(3+\sqrt{20}+\frac{1}{8\beta}\right)\sqrt{\sum_{j\ne i}P_j(T)}
     +\left(\frac{10}{\sqrt8}+\frac18\right)\sum_{j\ne i}\sqrt{Q_j(T)}.
\end{align*}
Since $10/\sqrt8+1/8<4$, this proves \eqref{eq:accumulated-loss-differences}.
\end{proof}

\Needspace{5\baselineskip}
\section{Adversarial regret}\label{app:adversarial}
\begin{theorem}[Adversarial regret]\label{thm:adversarial}
Fix integers $d\ge\max\{2,d_i\}$ and $d_i\ge1$, and $\beta\ge4$. Then, \algorithmref{alg:caoh} satisfies the following for every loss sequence in $[0,1]^{d_i}$ and integer $T\ge \beta^2\log d$:
\begin{equation}
 \Reg_i^T\le9\sqrt{T\log d}.
 \label{eq:adversarial}
\end{equation}
\end{theorem}
\begin{proof}
We first bound the squared inverse step sizes. By \cref{lem:soft-regret-decomposition,lem:kl-bounds},
\[
 |\hat R_i^t-\hat R_i^{t-1}|
 \le\frac1{2\beta_i^t}+(\log d_i)|\beta_i^t-\beta_i^{t-1}|.
\]
For every $t\ge1$, \eqref{eq:generic-rates} in \cref{lem:rates} gives $\Lambda_d|\beta_i^{t+1}-\beta_i^t|\le|\hat R_i^t-\hat R_i^{t-1}|$ and $\beta_i^{t+1}/\beta_i^t\le21/20<5/4$. Combining these bounds gives
\begin{align}
 \Lambda_d\bigl|(\beta_i^{t+1})^2-(\beta_i^t)^2\bigr|
 &\le(\beta_i^{t+1}+\beta_i^t)|\hat R_i^t-\hat R_i^{t-1}|\nonumber\\*
 &\le\frac94\left[\frac12+(\log d_i)\beta_i^t|\beta_i^t-\beta_i^{t-1}|\right]\nonumber\\*
 &\le\frac98+\frac{\Lambda_d}{4}\bigl|(\beta_i^t)^2-(\beta_i^{t-1})^2\bigr|.
 \label{eq:squared-inverse-step-recurrence}
\end{align}
The last inequality uses $\beta_i^t|\beta_i^t-\beta_i^{t-1}|\le|(\beta_i^t)^2-(\beta_i^{t-1})^2|$ and $9\log d_i\le\Lambda_d$.

We prove $|(\beta_i^{t+1})^2-(\beta_i^t)^2|\le3/(2\Lambda_d)$ for every $t\ge0$ by induction. At $t=0$, the difference is zero because $\beta_i^0=\beta_i^1=\beta$. For $t\ge1$, suppose $|(\beta_i^t)^2-(\beta_i^{t-1})^2|\le3/(2\Lambda_d)$. Substituting this hypothesis into \eqref{eq:squared-inverse-step-recurrence} gives
\[
 \Lambda_d\bigl|(\beta_i^{t+1})^2-(\beta_i^t)^2\bigr|
 \le\frac98+\frac{\Lambda_d}{4}\frac3{2\Lambda_d}
 =\frac32.
\]
Dividing by $\Lambda_d$ proves the bound at $t$ and completes the induction. For every $T\ge1$, it follows that
\begin{align}
 (\beta_i^T)^2
 &=\beta^2+\sum_{t=1}^{T-1}\bigl[(\beta_i^{t+1})^2-(\beta_i^t)^2\bigr]
 \le\beta^2+\frac{3(T-1)}{2\Lambda_d},\nonumber\\*
 (\hat R_i^T)^2
 &=\Lambda_d^2\bigl((\beta_i^{T+1})^2-\beta^2\bigr)\le\frac32\Lambda_dT.
 \label{eq:growth}
\end{align}
The first line telescopes from $\beta_i^1=\beta$ and bounds each signed increment by its absolute value. The second rearranges the update $(\beta_i^{T+1})^2=\beta^2+(\hat R_i^T/\Lambda_d)^2$ and applies the first line at $T+1$.
Using these bounds and $\sqrt{a+b}\le\sqrt a+\sqrt b$ in \cref{lem:regret-comparison}, we obtain
\begin{equation}
 \Reg_i^T\le \hat{R}_i^T+\beta_i^T\log d_i
 \le \beta\log d_i+\sqrt{\frac{3T}{2}}
 \left(\sqrt{\Lambda_d}+\frac{\log d_i}{\sqrt{\Lambda_d}}\right)
 \le9\sqrt{T\log d}.
 \label{eq:adv-proof}
\end{equation}
The second term is at most $8\sqrt{T\log d}$: using $\log d_i\le\log d$, its squared coefficient is at most $(3/2)(\Lambda_d/\log d+2+\log d/\Lambda_d)\le(3/2)(40+2+1/16)<64$. Also, $T\ge \beta^2\log d$ and $d_i\le d$ give $\beta\log d_i\le\sqrt{T\log d}$.
\end{proof}

\section{The failure of adaptation based on standard regret}\label{app:ordinary-regret}

In this section, we argue why using soft regret is critical for our algorithm.
To do so, we consider the following variant of \algorithmref{alg:caoh} that replaces the soft regret with the standard regret (while keeping other parts unchanged):
\begin{equation}
 \beta_i^{t+1}=\sqrt{\beta^2+(\Reg_i^t/\Lambda_d)^2},
 \qquad x_i^{t+1}=\softmax\!\left(-\frac{L_i^t+\ell_i^t}{\beta_i^{t+1}}\right),
 \qquad L_i^0=\ell_i^0=0.
 \label{eq:ordinary-update}
\end{equation}
We will show both theoretically and empirically that this variant can suffer linear regret.

\subsection{A linear regret lower bound}\label{app:ordinary-regret-lower-bound}
We construct a symmetric two-player game with four actions per player. Set $\nu=\Lambda_4^{-4}=[16(1+\log 4)]^{-4}$ and let $0<\kappa<1/4$. For either player, the rows of the loss matrix $G$ index the player's actions and the columns index the opponent's actions.
\begin{equation}
 G=\frac18\begin{pmatrix}
 4&4&4&4\\
 4+\kappa(2-\nu)&4+\kappa(2+7\nu)&4-\kappa(6+\nu)&4\\
 4-\kappa(1+2\nu)&4+\kappa(7-2\nu)&4+\kappa(6\nu-1)&4\\
 8&8&8&8
 \end{pmatrix}.
 \label{eq:ordinary-game}
\end{equation}
For $j\ne i$, player $i$'s loss vector is $Gx_j$. Write $x^\star=(5/8,1/8,1/4,0)^\top$ and $x_{2{:}3}=(x_2,x_3)^\top$ for the probabilities of actions 2 and 3 in a strategy $x$. Action 1 has loss $1/2$, action 4 has loss $1$, and the losses of actions 2 and 3 relative to action 1 satisfy
\begin{equation}\label{eq:ordinary-game-gaps}
 (Gx)_{2{:}3}-\tfrac12\one
 =\kappa\begin{pmatrix}\nu&-1\\1&\nu\end{pmatrix}
 \bigl(x_{2{:}3}-(1-x_4)x^\star_{2{:}3}\bigr).
\end{equation}
Since $\nu<1/8$, every entry in rows 2 and 3 of \eqref{eq:ordinary-game} is within $\kappa$ of $1/2$. Each expected loss is a convex combination of the entries in its row, so both loss gaps have absolute value at most $\kappa$. At $x=x^\star$, both gaps vanish, so $(x^\star,x^\star)$ is a Nash equilibrium.

\begin{theorem}[Linear regret with standard-regret adaptation]\phantomsection\label{thm:ordinary-linear}
There exist fixed $0<\kappa<1/4$ and $c>0$ such that the two-player game \eqref{eq:ordinary-game} with four actions per player and losses in $[0,1]$ has the following property. For every common base inverse step size $\beta>0$, self-play under \eqref{eq:ordinary-update} from uniform initialization satisfies $\Reg_i^T\ge cT$ for both players and all integers $T\ge1$.
\end{theorem}

\begin{proof}
Fix $\beta>0$. By symmetry and uniform initialization, the two players have the same strategy $x^t$, loss vector $\ell^t=Gx^t$, cumulative losses $L^t$, and regret $R^t$. To track how close their strategies are to $x^\star$, write $D^t=(L_2^t-L_1^t,L_3^t-L_1^t)^\top$ and set
\begin{equation}\label{eq:ordinary-u-definition}
 u^t=\frac{D^t}{\beta_i^{t+1}}-r^\star,\qquad
 r^\star=\left(\log\frac{x_1^\star}{x_a^\star}\right)_{a=2,3}
   =(\log 5,\log(5/2))^\top.
\end{equation}
The constants in the $O(\cdot)$ bounds below are independent of $t$, $\beta$, and sufficiently small $\kappa$.

\paragraph{Step 1. A positive lower bound on $\norm{u^t}$ gives linear regret.}
Suppose $\norm{u^t}\ge\delta$ for all $t$ and some fixed $\delta>0$. We show that, for sufficiently small $\kappa$, each step contributes a positive constant to regret against $x^\star$.

At a symmetric profile $(x,x)$, we subtract $1/2$ from every entry of $Gx$ and substitute the loss gaps in \eqref{eq:ordinary-game-gaps}. This gives
\begin{align}
 \ip{x-x^\star}{Gx}
 &=\kappa\nu\norm{x_{2{:}3}-(1-x_4)x^\star_{2{:}3}}^2
   +x_4\bigl(\tfrac12-\ip{x^\star_{2{:}3}}{(Gx)_{2{:}3}-\tfrac12\one}\bigr)\notag\\
 &\ge\kappa\nu\norm{x_{2{:}3}-(1-x_4)x^\star_{2{:}3}}^2+\tfrac14 x_4 .
 \label{eq:ordinary-comparator-regret}
\end{align}
The off-diagonal entries $1$ and $-1$ in \eqref{eq:ordinary-game-gaps} cancel in the quadratic term. The lower bound in \eqref{eq:ordinary-comparator-regret} uses $\bigl|\ip{x^\star_{2{:}3}}{(Gx)_{2{:}3}-\tfrac12\one}\bigr|\le 3\kappa/8$. In particular, $\ip{x-x^\star}{Gx}>0$ for $x\ne x^\star$, and $\ip{x^1-x^\star}{\ell^1}\ge1/16$ since $x^1_4=1/4$. Also, $x^\star$ gives mass at least $1/8$ to each of actions 1--3, so its cumulative loss exceeds the minimum by at least one eighth of the loss range among these actions. Hence
\begin{equation}\label{eq:ordinary-regret-floor}
 R^t=\sum_{k=1}^t\ip{x^k-x^\star}{\ell^k}+\ip{x^\star}{L^t}-\min_a L_a^t
 \ge \tfrac1{16}+\tfrac18\norm{D^t}_\infty .
\end{equation}
It follows that $\beta_i^{t+1}\ge 1/(16\Lambda_4)$. Writing $\Delta D^t=D^t-D^{t-1}$, the softmax update gives
\begin{equation}\label{eq:ordinary-conditional-play}
 \left(\log\frac{x_1^{t+1}}{x_a^{t+1}}\right)_{a=2,3}
 =r^\star+u^t+\frac{\Delta D^t}{\beta_i^{t+1}},
 \qquad
 \frac{x_4^{t+1}}{x_1^{t+1}}=e^{-(t+1)/(2\beta_i^{t+1})}.
\end{equation}
Since $\norm{\Delta D^t}_\infty\le\kappa$ and $\beta_i^{t+1}\ge1/(16\Lambda_4)$, we can choose $\kappa$ small enough that $\norm{\Delta D^t/\beta_i^{t+1}}<\delta/2$ for every $t$. Applying the reverse triangle inequality to \eqref{eq:ordinary-conditional-play} gives
\begin{equation}\label{eq:ordinary-log-ratio-separation}
 \left\lVert\left(\log\frac{x_1^{t+1}}{x_b^{t+1}}\right)_{b=2,3}-r^\star\right\rVert
 \ge \norm{u^t}-\norm{\Delta D^t/\beta_i^{t+1}}>\delta/2.
\end{equation}
The log ratios $(\log(x_1/x_b))_{b=2,3}$ are continuous at $x^\star$, where they equal $r^\star$. Hence, we can choose $\varepsilon>0$ such that $\norm{x-x^\star}<\varepsilon$ implies $\norm{(\log(x_1/x_b))_{b=2,3}-r^\star}\le\delta/2$. Hence, if $\norm{x^{t+1}-x^\star}<\varepsilon$ for some $t\ge1$, the log ratios of $x^{t+1}$ would contradict the strict lower bound $\delta/2$ in \eqref{eq:ordinary-log-ratio-separation}. Thus $\norm{x^{t+1}-x^\star}\ge\varepsilon$ for every $t\ge1$. Since $x^1$ differs from $x^\star$, we can also take $\varepsilon\le\norm{x^1-x^\star}$. The lower bound in \eqref{eq:ordinary-comparator-regret} has a positive minimum $c$ on $\{x\in\Delta_4:\norm{x-x^\star}\ge\varepsilon\}$, so $R^T\ge\sum_{t=1}^T\ip{x^t-x^\star}{\ell^t}\ge cT$.

\paragraph{Step 2. The recurrence of $u^t$.}
We study the linear term of the update near $u^t=0$ to show that $\norm{u^t}$ stays bounded away from zero. The linear approximation to the change in inverse step size uses the derivative $s_t=R^t/(\Lambda_4^2\beta_i^{t+1})$ of $\beta_i^{t+1}$ with respect to $R^t$ in \eqref{eq:ordinary-update}. We show that for some $a\in(0,1]$ and $\tau>0$, whenever $t\ge1$ and $\norm{u^t}\le a$,
\begin{equation}\label{eq:ordinary-local-recurrence}
 \frac{\beta_i^{t+2}}{\kappa}(u^{t+1}-u^t)
 =M(s_t)u^t+O(\norm{u^t}^2+\kappa),
\end{equation}
where the $2\times2$ matrix $M(s_t)$ satisfies $\operatorname{tr}M(s_t)\ge\tau$. Write $\Delta\beta_i^{t+2}=\beta_i^{t+2}-\beta_i^{t+1}$. Expanding $\Delta D^{t+1}$ using \eqref{eq:ordinary-u-definition} gives the exact update
\begin{equation}\label{eq:ordinary-incoming}
 \beta_i^{t+2}(u^{t+1}-u^t)=\Delta D^{t+1}-(r^\star+u^t)\Delta\beta_i^{t+2}.
\end{equation}

To identify $M(s_t)$ and bound the remainder in \eqref{eq:ordinary-local-recurrence}, we first express $\Delta\beta_i^{t+2}$ in terms of $\Delta D^{t+1}$. We will show $\Delta\beta_i^{t+2}=s_t\ip{x_{2{:}3}^{t+1}}{\Delta D^{t+1}}+O(\kappa^2)$ when $\norm{u^t}\le a$. Write $\Delta R^{t+1}=R^{t+1}-R^t$. Squaring the inverse step sizes in \eqref{eq:ordinary-update} and using $\beta_i^{t+1}\ge1/(16\Lambda_4)$ yields
\begin{equation*}
 \Delta\beta_i^{t+2}-s_t\Delta R^{t+1}
 =\frac{(\Delta R^{t+1})^2/\Lambda_4^2-(\Delta\beta_i^{t+2})^2}{2\beta_i^{t+1}}
 =O((\Delta R^{t+1})^2).
 \tag{\Cref{lem:inverse-step-lipschitz}}
\end{equation*}
Since $s_t\in[0,1/\Lambda_4]$ by \eqref{eq:ordinary-update} and \eqref{eq:ordinary-regret-floor}, it suffices to show $\Delta R^{t+1}=\ip{x_{2{:}3}^{t+1}}{\Delta D^{t+1}}+O(\kappa^2)$, which also gives $|\Delta R^{t+1}|=O(\kappa)$ by $\norm{\Delta D^{t+1}}_\infty\le\kappa$.
The regret increment subtracts the comparator loss increase $\min_b L_b^{t+1}-\min_b L_b^t$ from the expected loss in round $t+1$. To compute this difference, fix $j\in\{t,t+1\}$ with $\norm{u^j}\le a$. For~$k=2,3$, \eqref{eq:ordinary-u-definition} gives $D_k^j\ge\beta_i^{j+1}/2>\kappa$ for small $a$ and $\kappa$, since $r_k^\star+u_k^j\ge1/2$ and $\beta_i^{j+1}\ge1/(16\Lambda_4)$. As $|\Delta D_k^{t+1}|\le\kappa$ and $L_4^m-L_1^m=m/2>0$, action~1 attains the minimum at both $t$ and $t+1$. Hence
\begin{align}
 \Delta R^{t+1}
 &=\ip{x^{t+1}}{\ell^{t+1}}-\bigl(\min_b L_b^{t+1}-\min_b L_b^t\bigr)\notag\\*
 &=\ip{x^{t+1}}{\ell^{t+1}}-\ell_1^{t+1}
 =\ip{x_{2{:}3}^{t+1}}{\Delta D^{t+1}}+\tfrac12 x_4^{t+1}\ge-\kappa.
 \label{eq:ordinary-regret-increment}
\end{align}
To complete the regret expansion, we need $x_4^{t+1}=O(\kappa^2)$ in \eqref{eq:ordinary-regret-increment}. By the ratio $x_4^{t+1}/x_1^{t+1}$ in \eqref{eq:ordinary-conditional-play}, it suffices to show $\beta_i^{t+1}\le3(t+1)\kappa$. For the same $j$,
\begin{align*}
 \tfrac12\beta_i^{j+1}
 &\le D_2^j=\sum_{m=1}^j\Delta D_2^m\le j\kappa,\\*
 \beta_i^{t+1}
 &\le\beta_i^{j+1}+\kappa/\Lambda_4
 \le(2j+1/\Lambda_4)\kappa\le3(t+1)\kappa,\\*
 x_4^{t+1}
 &\le e^{-(t+1)/(2\beta_i^{t+1})}\le e^{-1/(6\kappa)}=O(\kappa^2).
\end{align*}
The bound $R^t\le R^j+\kappa$ holds for $j=t$ and follows from
\eqref{eq:ordinary-regret-increment} for $j=t+1$. Let $b$ be the inverse step
size as a function of regret in \eqref{eq:ordinary-update}. Since the regrets
are nonnegative by \eqref{eq:ordinary-regret-floor}, monotonicity and
\cref{lem:inverse-step-lipschitz} give
$b(R^t)\le b(R^j+\kappa)\le b(R^j)+\kappa/\Lambda_4$.
The remaining inequalities for $\beta_i^{t+1}$ use
$\beta_i^{j+1}\le2j\kappa$ from the first line, $j\le t+1$, and $\Lambda_4>1$.
Substituting $x_4^{t+1}=O(\kappa^2)$ into \eqref{eq:ordinary-regret-increment}
gives $|\Delta R^{t+1}|=O(\kappa)$ whenever
$\min\{\norm{u^t},\norm{u^{t+1}}\}\le a$. Taking $j=t$ yields the expansion
of $\Delta\beta_i^{t+2}$.

Now it remains to show that, for $\norm{u^t}\le a$,
$\Delta D^{t+1}/\kappa$ has a linear term in $u^t$ with remainder
$O(\norm{u^t}^2+\kappa)$. By \eqref{eq:ordinary-game-gaps}, we need to expand
$x_{2{:}3}^{t+1}-(1-x_4^{t+1})x^\star_{2{:}3}$.
Let $z=(\log(x_1^{t+1}/x_k^{t+1}))_{k=2,3}$, and write
$p_k(z)=e^{-z_k}/(1+e^{-z_2}+e^{-z_3})$ for the probability of action $k$
conditional on choosing among actions 1--3. Thus
$x_{2{:}3}^{t+1}=(1-x_4^{t+1})p(z)$ and
$p(r^\star)=x^\star_{2{:}3}$ since $x_4^\star=0$.
By \eqref{eq:ordinary-conditional-play}, $z-r^\star=u^t+O(\kappa)$.
Writing $Dp(r^\star)$ for the Jacobian of $p$ at $r^\star$, we obtain
\begin{align*}
 p(z)-p(r^\star)
 &=Dp(r^\star)(z-r^\star)+O(\norm{z-r^\star}^2)
 \tag{Taylor expansion of $p$ at $r^\star$}\\*
 &=Dp(r^\star)u^t+O(\norm{u^t}^2+\kappa),\\*
 \frac{\Delta D^{t+1}}{\kappa}
 &=(1-x_4^{t+1})\begin{pmatrix}\nu&-1\\1&\nu\end{pmatrix}
       [p(z)-p(r^\star)]\\*
 &=\underbrace{-\tfrac1{64}
   \begin{pmatrix}\nu&-1\\1&\nu\end{pmatrix}
   \begin{pmatrix}7&-2\\-2&12\end{pmatrix}}_{M(0)}u^t
   +O(\norm{u^t}^2+\kappa).
 \tag{$x_4^{t+1}=O(\kappa^2)$}
\end{align*}
The $(k,m)$ entry of $Dp(r^\star)$ is
$x_k^\star(x_m^\star-\one\{k=m\})$. Substituting
$x_2^\star=1/8$ and $x_3^\star=1/4$ gives the second matrix in $M(0)$,
including the factor $-1/64$. Also,
$x_{2{:}3}^{t+1}-x^\star_{2{:}3}
=(1-x_4^{t+1})[p(z)-p(r^\star)]-x_4^{t+1}x^\star_{2{:}3}
=O(\norm{u^t}+\kappa)$.

Substituting the expansions of $\Delta\beta_i^{t+2}$ and $\Delta D^{t+1}/\kappa$ into \eqref{eq:ordinary-incoming} gives
\begin{align*}
 \frac{\beta_i^{t+2}}{\kappa}(u^{t+1}-u^t)
 &=\bigl[I-s_t(r^\star+u^t)(x_{2{:}3}^{t+1})^\top\bigr]
     \frac{\Delta D^{t+1}}{\kappa}+O(\kappa)\\*
 &=\bigl[I-s_t r^\star(x^\star_{2{:}3})^\top\bigr]\frac{\Delta D^{t+1}}{\kappa}
     +O(\kappa)+O((\norm{u^t}+\kappa)^2)\\*
 &=\bigl[I-s_t r^\star(x^\star_{2{:}3})^\top\bigr]M(0)u^t
     +O(\norm{u^t}^2+\kappa).
 \tag{expanding $\Delta D^{t+1}/\kappa$}
\end{align*}
The second equality uses $(r^\star+u^t)(x_{2{:}3}^{t+1})^\top=r^\star(x^\star_{2{:}3})^\top+O(\norm{u^t}+\kappa)$. The matrix error multiplies $s_t\Delta D^{t+1}/\kappa=O(\norm{u^t}+\kappa)$, giving the error $O((\norm{u^t}+\kappa)^2)$. Thus \eqref{eq:ordinary-local-recurrence} holds with $M(s)=(I-s r^\star(x^\star_{2{:}3})^\top)M(0)$.

To show the positivity of $\operatorname{tr}M(s_t)$, we compute
\begin{align*}
 \operatorname{tr}M(s_t)
 &=\operatorname{tr}M(0)-s_t(x^\star_{2{:}3})^\top M(0)r^\star
 \tag{$\operatorname{tr}(vw^\top)=w^\top v$}\\*
 &=-\tfrac{19\nu}{64}+\tfrac{s_t\log 2}{32}
     +\tfrac{s_t\nu}{512}(3r_2^\star+22r_3^\star)
 \ge-\tfrac{19\nu}{64}+\tfrac{s_t\log2}{32}.
\end{align*}
The coefficient of $s_t$ in $\operatorname{tr}M(s_t)$ follows from $-(x^\star_{2{:}3})^\top M(0)=(16+3\nu,-16+22\nu)/512$ and $r_2^\star-r_3^\star=\log2$. To show that $s_t\log 2/32$ exceeds $19\nu/64$, we use $r_2^\star+u_2^t\ge\log5-a\ge1$ for small $a$. Hence
\begin{align*}
 s_t
 &\overset{\eqref{eq:ordinary-regret-floor}}{\ge}\frac{\norm{D^t}_\infty}{8\Lambda_4^2\beta_i^{t+1}}
 \overset{\eqref{eq:ordinary-u-definition}}{=}\frac{\norm{r^\star+u^t}_\infty}{8\Lambda_4^2}
 \ge\frac1{8\Lambda_4^2},\\*
 \operatorname{tr}M(s_t)
 &\ge-\tfrac{19\nu}{64}+\tfrac{\log2}{256\Lambda_4^2}
 =\frac{\Lambda_4^2\log2-76}{256\Lambda_4^4}
 =:\tau>0.
 \tag{$\nu=\Lambda_4^{-4}$, $\Lambda_4>32$}
\end{align*}

\Needspace{9\baselineskip}
\paragraph{Step 3. A positive lower bound on $\norm{u^t}$.}
By Step~1, it suffices to bound $\norm{u^t}$ away from zero uniformly in $\beta$. The trace in \eqref{eq:ordinary-local-recurrence} controls the leading change under the linear update. Write $M_t=M(s_t)$ and $B_t=I+h_tM_t$ for this update, with $h_t=\kappa/\beta_i^{t+2}$. Replacing $u^t$ by $B_tu^t$ with $M_t$ fixed sends $M_tu^t$ to $M_tB_tu^t=B_tM_tu^t$. Thus the area of the parallelogram with sides $u^t$ and $M_tu^t$ is multiplied by $|\det B_t|$. We show that, for fixed $c_0,C_0>0$, its area satisfies
\begin{equation}\label{eq:ordinary-area}
 A_t:=\bigl|\det[u^t,M_tu^t]\bigr|,\qquad
 c_0\norm{u^t}^2\le A_t\le C_0\norm{u^t}^2 \qquad(t\ge1).
\end{equation}
With \eqref{eq:ordinary-area}, it suffices to show that $A_t\ge c_0\rho^2$ at every $t$ for a small fixed $\rho>0$, since then $\norm{u^t}\ge\rho\sqrt{c_0/C_0}$. Indeed, if $A_{t+1}<c_0\rho^2\le A_t$ holds for some $t$, then $\norm{u^{t+1}}<\rho$ by \eqref{eq:ordinary-area}. Taking $\rho\le a/2$ gives $\norm{u^{t+1}}<a$, so the estimate $|\Delta R^{t+1}|=O(\kappa)$ after \eqref{eq:ordinary-regret-increment} applies. Using this estimate in \eqref{eq:ordinary-incoming} gives $\norm{u^{t+1}-u^t}=O(\kappa)$, hence $\norm{u^t}\le\rho+O(\kappa)\le a$ for sufficiently small $\kappa$. We can then use \eqref{eq:ordinary-local-recurrence} at $t$ to show $A_{t+1}>A_t$, a contradiction.

To prove \eqref{eq:ordinary-area}, we check that $v$ and $M(s)v$ are linearly independent for every unit $v\in\R^2$ and $0\le s\le1/\Lambda_4$. From Step~2, we have
\begin{align*}
 \det M(s)
 &=\bigl(1-s(x^\star_{2{:}3})^\top r^\star\bigr)\frac{5(1+\nu^2)}{256}
 >\frac1{64},\\*
 |\operatorname{tr}M(s)|&\le\nu+s<\frac1{16}.
\end{align*}
The first line uses $\det(I-sr^\star(x^\star_{2{:}3})^\top)=1-s(x^\star_{2{:}3})^\top r^\star$. The inequalities use $s,\nu<1/32$ and $r_2^\star,r_3^\star<2$. Thus the characteristic polynomial $\lambda^2-(\operatorname{tr}M(s))\lambda+\det M(s)$ has negative discriminant $(\operatorname{tr}M(s))^2-4\det M(s)$, so both eigenvalues of $M(s)$ are nonreal. If the unit vector $v$ and $M(s)v$ were linearly dependent, then $M(s)v=\lambda v$ for some real $\lambda$. This would give a real eigenvalue of $M(s)$, contradicting the negative discriminant. Hence $|\det[v,M(s)v]|$ is positive, and continuity on the compact set of $s\in[0,1/\Lambda_4]$ and unit $v$ gives $c_0\le|\det[v,M(s)v]|\le C_0$ for fixed $c_0,C_0>0$. For $u^t\ne0$, take $v=u^t/\norm{u^t}$. Bilinearity of the determinant gives $A_t=\norm{u^t}^2|\det[v,M(s_t)v]|$, proving \eqref{eq:ordinary-area}. For $u^t=0$, \eqref{eq:ordinary-area} holds with $A_t=0$.

Now we show that $\norm{u^t}\le a$ whenever $A_{t+1}<c_0\rho^2\le A_t$, so that \eqref{eq:ordinary-local-recurrence} applies at such $t$. For $0<\rho\le a/2$, \eqref{eq:ordinary-area} gives $\norm{u^{t+1}}<\rho\le a$. Then $|\Delta R^{t+1}|=O(\kappa)$ by the estimate after \eqref{eq:ordinary-regret-increment}, and $|\Delta\beta_i^{t+2}|=O(\kappa)$ by \cref{lem:inverse-step-lipschitz}. For $0<\kappa\le\rho^2$ and small $\rho$, \eqref{eq:ordinary-incoming} written in terms of $u^{t+1}$ gives
\[
 \norm{u^t}\le \norm{u^{t+1}}
   +\frac{\norm{\Delta D^{t+1}-(r^\star+u^{t+1})\Delta\beta_i^{t+2}}}{\beta_i^{t+1}}
   \le \rho+O(\kappa)\le 2\rho\le a .
\]
The numerator is $O(\kappa)$ because $\norm{\Delta D^{t+1}}\le\sqrt{2}\norm{\Delta D^{t+1}}_\infty\le\sqrt{2}\kappa$, $\norm{u^{t+1}}<\rho$, and $|\Delta\beta_i^{t+2}|=O(\kappa)$. By \eqref{eq:ordinary-regret-floor}, $\beta_i^{t+1}\ge1/(16\Lambda_4)$, so the fraction is $O(\kappa)$.

Now, we prove $A_t\ge c_0\rho^2$ for all $t\ge1$ by contradiction. Suppose there exists $t\ge1$ such that $A_{t+1}<c_0\rho^2\le A_t$ holds. At such $t$, \eqref{eq:ordinary-area} gives $\norm{u^{t+1}}<\rho$ and Step~2 gives $\operatorname{tr}M_t\ge\tau>0$ since $\norm{u^t}\le a$. The inequality $\operatorname{tr}M_t\ge\tau$ and $\det M_t>0$ give $\det B_t-1=h_t(\operatorname{tr}M_t+h_t\det M_t)\ge h_t\tau>0$.

Write $v=B_tu^t$ and $w=u^{t+1}$. With $M_t$ fixed, the linear update increases the area from $A_t$ to $\det(B_t)A_t=|\det[v,M_tv]|$. By \eqref{eq:ordinary-area}, $A_{t+1}=|\det[w,M_{t+1}w]|$. For a fixed $C>0$,
\begin{align*}
 A_{t+1}-A_t
 &=\bigl(\det(B_t)-1\bigr)A_t+\bigl[A_{t+1}-\det(B_t)A_t\bigr]\\*
 &\ge c_0\tau h_t\rho^2-\bigl|A_{t+1}-\det(B_t)A_t\bigr|,
 \tag{$A_t\ge c_0\rho^2$}\\*
 \bigl|A_{t+1}-\det(B_t)A_t\bigr|
 &\le \bigl|\det[w,M_{t+1}w]-\det[v,M_tv]\bigr|\\*
 &=\bigl|\det[w-v,M_tv]+\det[w,M_t(w-v)]+\det[w,(M_{t+1}-M_t)w]\bigr|\\*
 &\le C\rho\norm{w-v}+C\rho^2|s_{t+1}-s_t|.
\end{align*}
We expand $\det[w,M_{t+1}w]-\det[v,M_tv]$ by adding and subtracting $\det[w,M_tv]$ and $\det[w,M_tw]$. Linearity of the determinant in the first column gives $\det[w-v,M_tv]$ with $M_tv$ fixed. With $w$ fixed in the first column, linearity in the second column gives the terms with $M_tw-M_tv=M_t(w-v)$ and $M_{t+1}w-M_tw=(M_{t+1}-M_t)w$. The final inequality uses $|\det[x,y]|\le\norm{x}\norm{y}$, $\norm{v},\norm{w}=O(\rho)$, and $\norm{M_{t+1}-M_t}=O(|s_{t+1}-s_t|)$ since $M(s)$ is affine in $s$. Here $\norm{v}=O(\rho)$ follows from $\norm{u^t}\le2\rho$, boundedness of $M_t$, and $h_t=O(\kappa)$ by \eqref{eq:ordinary-regret-floor}.

It suffices to bound $|A_{t+1}-\det(B_t)A_t|$ by $O(h_t\rho^3)$. We bound $\norm{u^{t+1}-B_tu^t}$ using \eqref{eq:ordinary-local-recurrence} and $|s_{t+1}-s_t|$ using \eqref{eq:ordinary-regret-increment}. Step~2 gives $\norm{\Delta D^{t+1}}_\infty=O(\kappa\rho)$ and $x_4^{t+1}=O(\kappa^2)$, so
\begin{align*}
 \norm{u^{t+1}-B_tu^t}
 &=O(h_t\rho^2),
 \tag{$\norm{u^t}\le2\rho$, $\kappa\le\rho^2$}\\*
 |\Delta R^{t+1}|
 &\overset{\eqref{eq:ordinary-regret-increment}}{\le}
 \norm{\Delta D^{t+1}}_\infty+\tfrac12x_4^{t+1}=O(\kappa\rho),\\*
 |s_{t+1}-s_t|
 &=\frac{|\Delta R^{t+1}/\Lambda_4^2-s_t\Delta\beta_i^{t+2}|}{\beta_i^{t+2}}
 \le\frac{2|\Delta R^{t+1}|}{\Lambda_4^2\beta_i^{t+2}}
 =O(h_t\rho),
 \tag{\Cref{lem:inverse-step-lipschitz}}\\*
 A_{t+1}-A_t
 &\ge c_0\tau h_t\rho^2-C'h_t\rho^3>0.
\end{align*}
In the bound on $|s_{t+1}-s_t|$, the factor $2$ comes from $s_t\le1/\Lambda_4$ and $|\Delta\beta_i^{t+2}|\le|\Delta R^{t+1}|/\Lambda_4$. The last line holds for a fixed $C'>0$ and sufficiently small $\rho$, contradicting $A_{t+1}<c_0\rho^2\le A_t$.

Since $D^1=O(\kappa)$, \eqref{eq:ordinary-u-definition} and \eqref{eq:ordinary-regret-floor} give $u^1=-r^\star+O(\kappa)$, so $A_1>c_0\rho^2$ by \eqref{eq:ordinary-area} for small $\rho$. Thus $A_t\ge c_0\rho^2$ for all $t\ge1$ by induction, and \eqref{eq:ordinary-area} gives $\norm{u^t}\ge\sqrt{c_0/C_0}\,\rho$ independently of $\beta$ and $\kappa$. Step~1 gives $\Reg_i^T\ge cT$ for sufficiently small $\kappa$.
\end{proof}

\begin{figure}[!ht]
\centering
\definecolor{figureTwoStandard}{HTML}{440154}
\definecolor{figureTwoCAOH}{HTML}{169C82}
{\sffamily\fontsize{8.72pt}{10.5pt}\selectfont
 \textcolor{figureTwoStandard}{\rule[.55ex]{2em}{.65pt}}\enspace CAOH w/ standard regret\qquad
 \textcolor{figureTwoCAOH}{%
  \rule[.55ex]{.44em}{.65pt}\hspace{.28em}%
  \rule[.55ex]{.44em}{.65pt}\hspace{.28em}%
  \rule[.55ex]{.44em}{.65pt}}\enspace CAOH\par}
\smallskip
\begin{minipage}[t]{.43\linewidth}
  \centering
  \includegraphics[width=\linewidth]{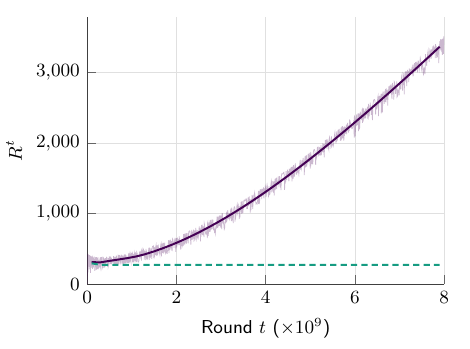}\par
  \hspace*{.133\linewidth}{\small (a)}
\end{minipage}\hspace{.06\linewidth}%
\begin{minipage}[t]{.43\linewidth}
  \centering
  \includegraphics[width=\linewidth]{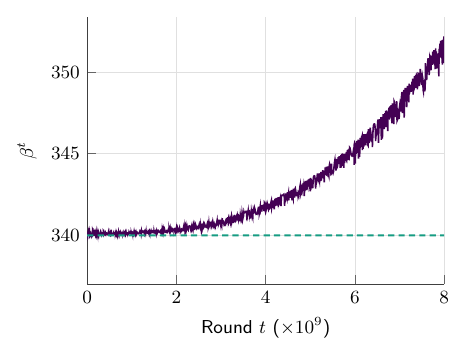}\par
  \hspace*{.133\linewidth}{\small (b)}
\end{minipage}
\caption[External regret and inverse step size in the two-player game]{%
\textbf{External regret and inverse step size in the two-player game.}
Solid purple and dashed teal curves represent the standard-regret variant in \eqref{eq:ordinary-update} and CAOH, respectively, both with $\beta=340$.
In (a), the thick curve averages the external regret $R^t$ over $2\times10^8$-round blocks, and smoothes the faint curve. The faint purple curve shows the variant's external regret $R^t$ at each $t$ shown.
In (b), the curves show the inverse step size $\beta^t=\beta_i^{t+1}$ for round $t+1$ at selected $t$.}
\label{fig:ordinary-regret-comparison}
\end{figure}

\Needspace{7\baselineskip}
\subsection{Numerical comparison}\label{app:ordinary-regret-experiment}

We use the loss gaps in \eqref{eq:ordinary-game-gaps} to compare CAOH with its standard-regret variant in \eqref{eq:ordinary-update}. We choose the common base inverse step size $\beta=340$ and use $x^\star_{2{:}3}=(1/20,1/3)^\top$, $\nu=5\times10^{-5}$, and $\kappa=0.24$ in \eqref{eq:ordinary-game-gaps}, chosen from a search over rounded values. Both players start uniformly and observe exact expected loss vectors. Their strategies coincide by symmetry, so we simulate one trajectory per update. We execute every round through $T=8\times10^9$ and plot the result in Figure~\ref{fig:ordinary-regret-comparison}.

\FloatBarrier
\section{Numerical experiment}\label{app:numerical-experiment}

In this section, we describe the games used in Figure~\ref{fig:dimension-sweeps}, which plots the maximum individual external regret $M_T=\max_i\Reg_i^T$. Our construction is inspired by \citet[Sections~3.4 and~6.3]{dagan_eisen_erez_koren_2026}, who use frequency doubling to generate fast OH oscillations and convert them into linear regret. We use a simpler two-player interaction that oscillates under OH at inverse step size $\beta=1/4$. Auxiliary players determine how long the loss matrix influences the two players' losses. In the dimensions $(n,d)$ tested, the limiting $M_T$ for OH increases with either $n$ or $d$.

\paragraph{Losses of players $X$ and $Y$.}
Players $X$ and $Y$ each have actions $1$ and $2$. The $k$ auxiliary players $Z_1,\ldots,Z_k$, which we call timers, each have action $1$ (\emph{wait}) and action $2$ (\emph{safe}). When at least one $Z_r$ chooses action $1$, $X$ and $Y$ share the loss matrix
\[
L=\begin{pmatrix}1&0\\0&1/2\end{pmatrix},
\]
where the row indexes the player's own action and the column indexes the other player's action. The game between $X$ and $Y$ with loss matrix $L$ has a symmetric Nash equilibrium in which both players use $(1/3,2/3)$. In the $n$-player games with additional timers, each of $X$ and $Y$ incurs loss $1/2$ for either of its actions when every $Z_r$ chooses action $2$.

For a pure action profile $a$, with $a_i$ the action of player $i$, the losses of $X$ and $Y$ are
\[
\bigl(\ell_X(a),\ell_Y(a)\bigr)
=\begin{cases}
(L_{a_X,a_Y},L_{a_Y,a_X}),& a_{Z_r}=1\text{ for some }r\in[k],\\
(1/2,1/2),& a_{Z_r}=2\text{ for all }r\in[k].
\end{cases}
\]

By symmetry, we have $x_X^t=x_Y^t$ for every $t$. Write $p_t=x_{X,1}^t=x_{Y,1}^t$ and let $q_t=1-\prod_{r=1}^k(1-x_{Z_r,1}^t)$ be the probability that at least one $Z_r$ chooses action $1$.

For $i\in\{X,Y\}$, we have $\ell_{i,1}^t-\ell_{i,2}^t=q_t(p_t-(1-p_t)/2)=\tfrac32q_t(p_t-1/3)$. Using the equilibrium as a fixed comparator gives
\begin{equation}\label{eq:experiment-equilibrium-regret}
\Reg_X^T=\Reg_Y^T\ge\sum_{t=1}^T(p_t-1/3)(\ell_{X,1}^t-\ell_{X,2}^t)
=\frac32\sum_{t=1}^Tq_t(p_t-1/3)^2.
\end{equation}

In our constructions, the timers keep $q_t\ge5/6$ over an initial interval. The next lemma shows that on this interval, OH makes $p_t$ alternate across $1/3$ while staying bounded away from it. Each round in the interval therefore contributes at least a positive constant to the regret lower bound in~\eqref{eq:experiment-equilibrium-regret}.

\begin{lemma}[Oscillation under Optimistic Hedge]\label{lem:experiment-oscillation}
Suppose $X$ and $Y$ run OH from uniform initialization with $\beta=1/4$. There is an absolute constant $\delta>0$ such that, for every integer $N\ge1$ with $q_t\ge5/6$ for $1\le t\le N$,
\[
 (-1)^{t-1}(p_t-1/3)\ge\delta\quad(1\le t\le N),
 \qquad
 \Reg_X^T=\Reg_Y^T\ge\tfrac54\delta^2 N\quad(T\ge N).
\]
\end{lemma}

\begin{proof}
Fix $N$ as in the statement. By~\eqref{eq:experiment-equilibrium-regret}, it suffices to prove that $p_t-1/3$ alternates in sign and stays bounded away from zero for $1\le t\le N$. Set $y_t=\log(2p_t/(1-p_t))$. Then, $p_t=e^{y_t}/(2+e^{y_t})$ and $y_t$ has the same sign as $p_t-1/3$. We will prove $(-1)^{t-1}y_t\ge\varepsilon$ for an absolute constant $\varepsilon>0$.

For $t\ge1$, write $h_t=(\ell_{X,1}^t-\ell_{X,2}^t)/\beta=6q_t(p_t-1/3)$, with $h_0=0$, and set $s_t=\log 2-\sum_{j=1}^{t-1}h_j$. The OH update with prediction $\ell_X^{t-1}$ gives
\begin{equation}\label{eq:experiment-odds-update}
 y_t=s_t-h_{t-1},\qquad s_{t+1}=s_t-h_t,\qquad y_{t+1}=s_t-2h_t
 \qquad(t\ge1).
\end{equation}
We claim that an absolute constant $\varepsilon\in(0,\log 2)$ satisfies, for $1\le t\le N$,
\begin{equation}\label{eq:experiment-oscillation-bounds}
 -3\le s_t\le5,\qquad
 \begin{cases}
  2h_t\ge\min\{y_t,5\}+\varepsilon,&\text{if }y_t\ge\varepsilon,\\
  -2h_t\ge\min\{-y_t,3\}+\varepsilon,&\text{if }y_t\le-\varepsilon.
 \end{cases}
\end{equation}
Since $s_t=y_t+h_{t-1}$ and $y_{t+1}=s_t-2h_t$,~\eqref{eq:experiment-oscillation-bounds} gives, for $1\le t<N$,
\[
 \begin{cases}
  y_{t+1}\le\min\{y_t,5\}-2h_t\le-\varepsilon,
   &\text{if }y_t\ge\varepsilon,\ h_{t-1}\le0,\\
  y_{t+1}\ge-\min\{-y_t,3\}-2h_t\ge\varepsilon,
   &\text{if }y_t\le-\varepsilon,\ h_{t-1}\ge0.
 \end{cases}
\]
Since $q_t\ge5/6$, $h_t$ has the same sign as $y_t$. In each case, $(y_{t+1},h_t)$ therefore satisfies the conditions of the other case at round $t+1$. Starting from $y_1=\log 2>\varepsilon$ and $h_0=0$, induction yields $(-1)^{t-1}y_t\ge\varepsilon$. As $y\mapsto e^y/(2+e^y)$ is strictly increasing, taking $\delta=\min_{\sigma\in\{-1,1\}}|e^{\sigma\varepsilon}/(2+e^{\sigma\varepsilon})-1/3|>0$ gives the asserted bound on $p_t-1/3$. For $T\ge N$,~\eqref{eq:experiment-equilibrium-regret} then gives
\[
 \Reg_X^T=\Reg_Y^T
 \ge\frac32\sum_{t=1}^N q_t(p_t-1/3)^2
 \ge\frac54\delta^2 N,
\]
where the first inequality drops the nonnegative terms with $t>N$.

It remains to prove~\eqref{eq:experiment-oscillation-bounds}. To bound $s_t$, we first verify $|h_t|\le|y_t|+1$. Since $p\mapsto\log(2p/(1-p))$ vanishes at $p=1/3$ and has derivative $1/(p(1-p))\ge4$, we have $|y_t|\ge4|p_t-1/3|$. Also $|h_t|\le6|p_t-1/3|$ since $q_t\le1$. If $p_t\le5/6$, subtracting the former from the latter gives $|h_t|-|y_t|\le(6-4)|p_t-1/3|\le1$. If $p_t>5/6$, the function $p\mapsto6(p-1/3)-\log(2p/(1-p))$ is decreasing on $[5/6,1)$ and is at most $1$ at $5/6$, so $|h_t|-|y_t|\le1$ in this case as well. Now $h_{t-1}\in[-2,4]$ and $h_t$ has the same sign as $y_t$, so~\eqref{eq:experiment-odds-update} gives $s_{t+1}=h_{t-1}+y_t-h_t$ and
\[
 \begin{cases}
  -3\le h_{t-1}-1\le s_{t+1}\le s_t,&\text{if }y_t\ge0,\\
  s_t\le s_{t+1}\le h_{t-1}+1\le5,&\text{if }y_t\le0.
 \end{cases}
\]
Together with $s_1=\log 2$, these inequalities prove $s_t\in[-3,5]$ by induction.

For the two bounds on $h_t$ in~\eqref{eq:experiment-oscillation-bounds}, use $q_t\ge5/6$ and $p_t=e^{y_t}/(2+e^{y_t})$. For $a\ge0$, substitution into $h_t=6q_t(p_t-1/3)$ gives
\[
 \begin{cases}
  h_t\ge\underbrace{\dfrac{10}{3}\dfrac{e^a-1}{e^a+2}}_{F(a)},&\text{if }y_t=a,\\[1ex]
  -h_t\ge\underbrace{\dfrac{10}{3}\dfrac{e^a-1}{1+2e^a}}_{H(a)},&\text{if }y_t=-a.
 \end{cases}
\]
It therefore suffices to show $2F(a)\ge\min\{a,5\}+\varepsilon$ and $2H(a)\ge\min\{a,3\}+\varepsilon$ for all $a\ge\varepsilon$. Differentiation shows that $F$ and $H$ are increasing on $[0,\infty)$ and $H$ is concave; their formulas also give $F\ge H$. Moreover, $H(0)=0$, $H'(0)=10/9$, and $e^3>16$ gives $2H(3)>3$ and $2F(3)>5$. Hence we can choose $\varepsilon\in(0,\log 2)$ small enough that $H(\varepsilon)\ge\varepsilon$, $2H(3)\ge3+\varepsilon$, and $2F(3)\ge5+\varepsilon$. On $[\varepsilon,3]$, the concave function $a\mapsto2H(a)-a$ is at least $\varepsilon$ at both endpoints, so $2F(a)\ge2H(a)\ge a+\varepsilon$ throughout this interval. For $a\ge3$, monotonicity gives $2H(a)\ge3+\varepsilon$ and $2F(a)\ge5+\varepsilon$. These estimates prove the two bounds on $h_t$ in~\eqref{eq:experiment-oscillation-bounds}.
\end{proof}

\paragraph{Increasing the number of players.}
To observe how regret can grow exponentially with the number of players, we vary the number $k$ of timers. Counting the timers and players $X$ and $Y$ gives $n=k+2$.
Each timer's safe action has zero loss, while its wait action is penalized only when every other timer chooses safe:
\[
\ell_{Z_r}(1,a_{-Z_r})
=\frac12\one\{a_{Z_j}=2\text{ for all }j\in[k]\setminus\{r\}\}.
\]
The timers start uniformly and, by symmetry, use the same wait probability $w_t=x_{Z_r,1}^t$ in every round. Their expected wait loss is $g_t=\tfrac12(1-w_t)^{k-1}$, and the matrix losses of $X$ and $Y$ have weight $q_t=1-(1-w_t)^k$. In particular, $g_1=2^{-k}$ and $q_1=1-2^{-k}$.

Fix $\beta=1/4$ and set $g_0=0$. If $w_s\ge1/3$ for all $1\le s<t$, then $g_s\le(2/3)^{k-1}/2$ on those rounds and the OH update gives
\[
\log\frac{1-w_t}{w_t}
=\frac{\sum_{s=1}^{t-1}g_s+g_{t-1}}{\beta}
\le\frac{t}{2\beta}(2/3)^{k-1}.
\]
The right-hand side is at most $\log 2$ for $t\le T_k=\lfloor 2\beta\log 2\,(3/2)^{k-1}\rfloor$. Induction from $w_1=1/2$ therefore gives $w_t\ge1/3$ and $q_t\ge1-(2/3)^k$ through $T_k$.
For sufficiently large $k$, $1-(2/3)^k\ge5/6$. Lemma~\ref{lem:experiment-oscillation} therefore gives $\Reg_X^T=\Reg_Y^T\ge\tfrac54\delta^2 T_k=\Omega((3/2)^k)$ for every $T\ge T_k$.

\paragraph{Increasing the number of actions.}
To observe how regret can grow exponentially with the number of actions, we keep four timers and add players $U,P,Q$, giving $n=9$. Player $U$ has $d$ zero-loss actions and stays uniform. For both $P$ and $Q$, action $1$ is wait and action $2$ has zero loss. Set
\begin{align*}
\ell_P(1,a_{-P})&=\tfrac12\one\{a_U=1\},\\
\ell_Q(1,a_{-Q})&=\tfrac12\one\{a_P=1\},\\
\ell_{Z_r}(1,a_{-Z_r})
&=\tfrac12\one\{a_Q=1,\ a_{Z_j}=2\text{ for all }j\in[k]\setminus\{r\}\}.
\end{align*}
By Lemma~\ref{lem:experiment-oscillation}, it suffices to show $q_t\ge5/6$ for $1\le t\le T_d=\lfloor c2^d\rfloor$, for some absolute constant $c>0$ and all sufficiently large $d$. By symmetry, the four timers have the same wait probability $w_t=x_{Z_r,1}^t$, so $q_t=1-(1-w_t)^4$. If $w_t\ge3/8$, then $q_t\ge1-(5/8)^4>5/6$.

Write $u_t=x_{Q,1}^t$ for $Q$'s wait probability. We show that $\sum_{s=2}^{T_d}u_s\le3/8$ suffices for $w_t\ge3/8$ for every $1\le t\le T_d$. Each timer's expected wait loss is $g_t=\tfrac12u_t(1-w_t)^3$, with $g_1=1/32$ from uniform initialization. If $w_s\ge3/8$ for $1\le s<t$, then $g_s\le u_s/8$ for $2\le s<t$. For $3\le t\le T_d$, the OH update at $\beta=1/4$ gives
\[
 \log\frac{1-w_t}{w_t}
 =4\left(g_1+\sum_{s=2}^{t-1}g_s+g_{t-1}\right)
 \le \frac18+\sum_{s=2}^{T_d}u_s\le\frac12.
\]
The log odds in rounds $1$ and $2$ are $0$ and $8g_1=1/4$, respectively. Since $r\mapsto(1+e^r)^{-1}$ is decreasing and $1/4$-Lipschitz, log odds at most $1/2$ give $w_t\ge(1+e^{1/2})^{-1}\ge3/8$, completing the induction.

It remains to show $\sum_{t=2}^{T_d}u_t\le3/8$. Write $v_t=x_{P,1}^t$ for $P$'s wait probability. Since $U$ stays uniform, $P$'s expected wait loss is $1/(2d)$, so OH gives $v_1=1/2$ and $v_t=(1+e^{2t/d})^{-1}$ for $t\ge2$. The expected wait loss of $Q$ is $v_t/2$, and $v_t$ stays bounded away from zero for the first $d$ rounds. Thus $Q$ accumulates order $d$ loss on its wait action over those rounds. The OH update gives, for $t\ge2$,
\begin{equation}\label{eq:experiment-Q-odds}
 \log\frac{1-u_t}{u_t}
 =2\left(\sum_{s=1}^{t-1}v_s+v_{t-1}\right)
 =d\log\frac{2}{1+e^{-2t/d}}+O(1).
\end{equation}
The last equality compares $\sum_{s=1}^{t-1}v_s$ with the integral of $(1+e^{2s/d})^{-1}$ over $[0,t]$. The monotonicity of $s\mapsto(1+e^{2s/d})^{-1}$ bounds the difference by $O(1)$, and $2v_{t-1}\le1$. Since the right-hand side of~\eqref{eq:experiment-Q-odds} is $d\log2+O(1)$ when $t>d^2$, $u_t=\Theta(2^{-d})$ holds for any $t>d^2$. Hence $\sum_{t=d^2+1}^{T_d}u_t=O(c)$.

For $\sum_{t=1}^{d^2}u_t$, we bound the terms with $t\ge2$, since the timer update accounts for $u_1=1/2$ through $g_1$. The first equality in~\eqref{eq:experiment-Q-odds} gives $u_t\to(1+e^t)^{-1}$ as $d\to\infty$ for each fixed $t\ge2$, since $v_s\to1/2$ for every fixed $s$. Also $2v_s\ge\gamma:=2/(1+e^2)$ holds for $1\le s\le d$, so we have $u_t\le e^{-\gamma\min\{t-1,d\}}$. Dominated convergence for $2\le t\le d$ and the bound $d^2e^{-\gamma d}$ on the sum over $d<t\le d^2$ give
\[
 \lim_{d\to\infty}\sum_{t=2}^{d^2}u_t
 =\sum_{t=2}^\infty\frac1{1+e^t}
 \le\sum_{t=2}^\infty e^{-t}<\frac14.
\]
Combining the sums over $2\le t\le d^2$ and $d^2<t\le T_d$ gives $\sum_{t=2}^{T_d}u_t\le3/8$ for a sufficiently small fixed $c>0$ and all sufficiently large $d$. Thus $w_t\ge3/8$ and $q_t\ge5/6$ through round $T_d$. Lemma~\ref{lem:experiment-oscillation} now gives $\Reg_X^T=\Reg_Y^T\ge\tfrac54\delta^2 T_d=\Omega(2^d)$ for every $T\ge T_d$.

\end{document}